\documentclass[%
 reprint,
groupedaddress,
nofootinbib,
 amsmath,amssymb,
 aps,
prd,
]{revtex4-1}

\usepackage{graphicx}
\usepackage{dcolumn}
\usepackage[utf8x]{inputenc}
\usepackage{bm}
\usepackage[colorlinks=true, linkcolor = red, citecolor = blue]{hyperref}

\begin{document}

\preprint{APS/123-QED}

\title{Tracker phantom field and a cosmological constant: dynamics of a composite dark energy model}

\author{Francisco X. Linares Cede\~no}  
  \email{linares@mctp.mx}
  \affiliation{%
 Mesoamerican Centre for Theoretical Physics, Universidad Autónoma de Chiapas, Carretera Zapata Km 4, Real del Bosque (Terán), 29040, Tuxla Gutierrez, Chiapas, México}
 
 \author{Nandan Roy}  
  \email{nandan.roy@mahidol.ac.th}
  \affiliation{%
 Centre for Theoretical Physics \& Natural Philosophy ``Nakhonsawan Studiorum for Advanced Studies", Mahidol University, Nakhonsawan Campus, Phayuha Khiri, Nakhonsawan 60130, Thailand}

\author{L. Arturo Ure\~na-L\'opez}
\email{lurena@ugto.mx}
\affiliation{%
  Departamento de F\'isica, DCI, Campus Le\'on, Universidad de Guanajuato, 37150, Le\'on, Guanajuato, M\'exico
}%

\date{\today}

\begin{abstract}
In this work, we study tracker phantom dark energy models with a general parameterization of the scalar potentials. Our analysis also considers the scenario of having both phantom field and the cosmological constant as the dark energy components. A detailed statistical analysis with current cosmological observations shows an increase in the value of the Hubble parameter due to the presence of phantom dark energy but it can not alleviate the Hubble tension completely. Our results using Bayesian methods suggests a decisive evidence in favor of a phantom field over a positive cosmological constant, although the possibility of a negative cosmological constant cannot be ruled out hidden in the dark sector.
\end{abstract}

\maketitle


\section{\label{sec:introduction}Introduction}

Over the years, different CMB experiments like WMAP~\cite{Hinshaw:2012aka} and Planck satellites~\cite{Aghanim:2018eyx, Akrami:2018vks} have constrained the standard $\Lambda$CDM model with unprecedented accuracy, and has made it the best observationally consistent model of the accelerating Universe. This enhancement of our ability to constrain the cosmological parameters with greater accuracy, has of lately evidenced a statistically significant tension in the estimation of $H_0$ between observations from the early Universe like CMB and BAO, and from observations from the late time Universe~\cite{Verde:2019ivm}. 

CMB Planck data~\cite{Aghanim:2018eyx} together with BAO~\cite{Alam_2017,Beutler:2011hx}, BBN~\cite{Alam:2020sor}, and DES~\cite{Troxel:2017xyo,Abbott:2017wau,Krause:2017ekm} have constraint the Hubble parameter to be $H_0 \sim (67.0 - 68.5)$km/s/Mpc. On the other hand, cosmic distance ladder and time delay measurement like those reported by SH0ES~\cite{Riess:2019cxk} and H0LiCOW~\cite{Wong:2019kwg} collaborations have reported $H_0 = (74.03 \pm 1.42)$km/s/Mpc and $H_0 = (73.3^{+1.7} _{-1.8})$km/s/Mpc respectively by observing the local Universe. In the beginning, there was speculation that this tension may have a systematic origin, but the persistence and increasing of such tension over the years (currently around 4.4$\sigma$), strongly suggests cosmologists should think about possibilities beyond $\Lambda$CDM. For a short update on the Hubble tension see~\cite{DiValentino:2020zio}, and for a detailed and comprehensive review see~\cite{DiValentino:2021izs}.

One of the proposed solutions to the Hubble tension, is the departure of the dark energy (DE) equation of state (EoS) from that of a cosmological constant $w_{DE} = -1$ to a phantom one $w_{DE} \leq -1$~\cite{Alestas:2020mvb,DiValentino:2020naf,Vagnozzi:2019ezj,DiValentino:2019dzu}. A phantom-like EoS of the DE can generate extra acceleration of the Universe compared to the cosmological constant, resulting in an increment of the value of the $H_0$. Generally these models can alleviate the  Hubble tension within $2 \sigma$.

Given the above motivation, here we make a revision of phantom models with scalar fields. Although scalar fields are widely used as alternatives to the cosmological constant, they suffer from the coincidence and fine-tuning problems. A probable way out for these models to alleviate these problems, is by considering the case of \textit{tracker solutions}~\cite{PhysRevD.37.3406,Steinhardt:1999nw}. In these solutions the scalar field energy density tracks the background dominating energy density, and behave as an attractor-like solution for a wide range of initial conditions. Recently, existence of a general class of tracker solution using a general parameterization of the scalar field potentials for quintessence models has been reported in~\cite{Urena-Lopez:2020npg}. These general tracker solutions not only track the background, but can also give us a late time behavior of the Universe consistent with observations. 

Unlike the quintessence models, the general tracking behavior of the phantom models has not received enough attention. Some studies have been done to study the tracking behavior of the phantom fields but for very specific cases~\cite{Chiba:2005tj,Saridakis:2009pj,Hao:2003th,Kujat:2006vj}. In this work, we study the tracking behavior of the phantom scalar field models for the same general parameterization used in~\cite{Urena-Lopez:2020npg,Roy:2018nce}, and show that it is possible to write down a general tracking condition for the phantom field and construct the corresponding solutions for a large class of potentials.

We shall also consider a scenario in which the DE sector consists of both the cosmological constant and the phantom field. It is customary to neglect the cosmological constant in alternative DE models, but in the case of scalar fields a constant potential term does not affect the field dynamics but only the density contribution to the Friedmann equation. Taking advantage of this, we will evaluate whether observations indicate any preference for the cosmological constant alone or for a composite model with more internal complexity. In doing so, we do not exclude beforehand the possibility of a negative cosmological constant, which has been recently considered in~\cite{Visinelli:2019qqu,Calderon:2020hoc,2020PhRvD.101h3507Y,2020PhRvD.101f3528A}.

The paper is organized in the following way. Section~\ref{sec:mathematical} deals with the construction of the dynamical systems for both the background and the perturbation equations of motion by using the hyperbolic polar transformations. In Sec.~\ref{sec:phantom} we discuss the existence of different types of solutions and the general condition for the tracking behavior using a parameterization of the scalar field potentials. The numerical evaluations of the phantom models are studied in Sec.~\ref{sec:tracker}. In Sec.~\ref{sec:numerical}, constraints on the cosmological parameters are given and Bayesian model comparison has been done. The conclusion and summary of the analysis are given in the Section~\ref{sec:discussion}.

\section{\label{sec:mathematical}Mathematical background}
The equation of motions for the phantom scalar field are revised here, following the same formalism for other scalar field models in~\cite{Urena-Lopez:2020npg,Roy:2018nce,Urena-Lopez:2015odd,Urena-Lopez:2015gur}, but with some necessary changes to take into account the phantom nature of the field. As mentioned above, the field dynamics is described for both the background and the linear perturbations, with the participation of both the phantom field and a cosmological constant.

\subsection{Phantom background evolution}
We consider a spatially flat, homogeneous and isotropic Universe described by the FRW metric filled with barotropic fluids and a phantom scalar field. The Einstein field equations together with the wave equation of the above mentioned Universe are,
\begin{subequations}
\label{eq:fri}
  \begin{eqnarray}
    H^2 = \frac{\kappa^2}{3} \left( \sum_j \rho_j +
      \rho_\phi \right) \, , \quad  \dot{\rho}_j = - 3 H (\rho_j + p_j ) \, \label{eq:fri_a} \\
    \dot{H} = - \frac{\kappa^2}{2} \left[ \sum_j (\rho_j +
      p_j ) + (\rho_\phi + p_\phi) \right] \, , \label{eq:fri_b} \\
    \ddot{\phi} = -3 H \dot{\phi} + \partial_\phi V(\phi) \, , \label{eq:1d}
  \end{eqnarray}
\end{subequations}
where $\kappa^2 = 8\pi G$, $\rho_j$ and $p_j$ are respectively the energy and pressure density of ordinary matter, a dot denotes derivative with respect to cosmic time $t$, and $H = \dot{a}/a$ is the Hubble parameter, with $a$ the scale factor of the Universe. 

The index $j$ runs over all the matter species in the Universe apart from the scalar field (e.g. photons, baryons, etc.), and the perfect fluids are related through the barotropic relation $p_j = (\gamma_j -1) \rho_j$. The barotropic equation of state (EoS) takes the usual values of $\gamma_j=4/3$ for a relativistic species, $\gamma_j=1$ for a nonrelativistic one, and $\gamma_j = 0$ for a cosmological constant. 

Given our interest to include a cosmological constant in our analysis, we note that the phantom potential can also be written in the form $V(\phi) = V_0 + V_1(\phi)$, where $V_0$ is simply a constant term and all the field dependencies in the potential are encoded in the term $V_1$. The expressions for the phantom energy density and pressure are, respectively, $\rho_\phi = -(1/2)\dot{\phi}^2+V_1(\phi) + V_0$ and $p_\phi = -(1/2)\dot{\phi}^2-V_1(\phi) - V_0$. Notice that the dynamics of the phantom field is not modified by the introduction of the constant term $V_0$ in the potential (see Eq.~\eqref{eq:1d}), but the latter only appears in the equations of motion for the Hubble parameter~\eqref{eq:fri_a} as an extra cosmological constant. 

Under this freedom to include a constant term in the phantom potential, we will refer to $\rho_\Lambda$ as the effective density that contains all possible constant terms in the total density, and likewise for the corresponding pressure which satisfies the relation $p_\Lambda = - \rho_\Lambda$. In line with this, and for simplicity in the notation, hereafter we make the change $V_1(\phi) \to V(\phi)$.

To ease the numerical solution of the phantom equation of motion, and inspired by the case of the quintessence field~\cite{Urena-Lopez:2015gur,Roy:2018nce,Urena-Lopez:2020npg}, we define a new set of hyperbolic \emph{polar} coordinates in the following form   
\begin{subequations}
\label{eq:hyper}
\begin{gather}
\frac{\kappa \dot{\phi}}{\sqrt{6} H}  \equiv  \Omega^{1/2}_\phi \sinh(\theta/2) \ \label{eq:hyper_a}, \quad \;
 \frac{\kappa V^{1/2}}{\sqrt{3} H} \equiv \Omega^{1/2}_\phi \cosh(\theta/2) \, , \\
y_1  \equiv -2\ \sqrt{2} \, \frac{\partial_{\phi}V^{1/2}}{H} \, , \quad y_2 \equiv - 4\sqrt{3} \frac{\partial^2_\phi V_{\phi}^{1/2}}{\kappa H} \, ,
\end{gather}
\end{subequations}
with which the Klein-Gordon equation~\eqref{eq:1d} is written as the following dynamical system,
\begin{subequations}
\label{eq:new4}
  \begin{eqnarray}
  \theta^\prime &=& -3 \sinh \theta - y_1 \label{eq:new4a} \, , \\
  y^\prime_1 &=& \frac{3}{2}\gamma_{tot} y_1 + \Omega_{\phi}^{1/2} \sinh(\theta /2) y_2 \, , \label{eq:new4c} \\
    \Omega^\prime_\phi &=& 3 (\gamma_{tot} - \gamma_\phi) \Omega_\phi \label{eq:new4b} \, .
\end{eqnarray}
\end{subequations}

The prime denotes derivative with respect to the number of $e$-foldings $N \equiv \ln (a/a_i)$, with $a_i$ the initial value of the scale factor. Here, $\gamma_{tot} = (p_{tot} + \rho_{tot})/\rho_{tot}$ is the total EoS written in terms of the total pressure $p_{tot}$ and total density  $\rho_{tot}$ of all the matter species. In particular, the EoS parameter of the phantom field can be written as $\gamma_\phi = (p_\phi + \rho_\phi)/\rho_\phi = 1 - \cosh \theta$.

A note is in turn. In the new variables~\eqref{eq:hyper} we assumed that $\Omega_\phi$ is positive definite, and in consequence so is the phantom density, $\rho_\phi = 3H^2 \Omega_\phi/\kappa^2 >0$. This is not necessarily the case of phantom fields, as for certain cases the energy density can be negative. However, we will consider initial conditions for a radiation dominated Universe, and then $\Omega_\phi \to 0^+$ at early times, which assures that $\Omega_\phi$ will be positive definite for the rest of the evolution.

\subsection{Phantom linear density perturbations}
Now, we are going to consider linear perturbations around the background values of the FRW line element (in the synchronous gauge),
\begin{equation}
ds^2 = -dt^2+a^2(t)(\delta_{ij}+h_{ij})dx^idx^j\, , \label{eq:mfpert}
\end{equation}
as well as for the scalar field in the form $\phi(\vec{x},t) = \phi(t)+\varphi(\vec{x},t)$. Here, $h_{ij}$ and $\varphi$ are the metric and scalar field perturbations, respectively. The linearized KG equation for the phantom field, for a Fourier mode $\varphi(k,t)$, reads~\cite{Ratra:1990me,Ferreira:1997au,Ferreira:1997hj,Perrotta:1998vf}:
\begin{equation}
  \ddot{\varphi}  = - 3H \dot{\varphi} - \left[\frac{k^2}{a^2} - \frac{\partial^2 V(\phi)}{\partial \phi^2}\right] \varphi - \frac{1}{2} \dot{\phi} \dot{\bar{h}} \, , \label{eq:perts}
\end{equation}
where $\bar{h}$ is the trace of the spatial part of the metric perturbation, and $k$ is its comoving wave number.

Again, the perturbed KG equation~\eqref{eq:perts} can be transformed into a dynamical system by using the following change of variables~\cite{Urena-Lopez:2015gur,Cedeno:2017sou},
\begin{subequations}
\label{eq:linearvars}
\begin{eqnarray}
\sqrt{\frac{2}{3}} \frac{\kappa \dot{\varphi}}{H} = -\Omega^{1/2}_{\phi} e^{\beta} \cosh(\vartheta/2) \, , \\ 
\frac{\kappa y_1 \varphi}{\sqrt{6}} = -\Omega^{1/2}_{\phi} e^{\beta} \sinh(\vartheta/2) \, , \label{eq:22a}
\end{eqnarray}
\end{subequations}
where $\beta$ and $\vartheta$ are the new variables introduced related to the evolution of the scalar field perturbation. With another set of variables defined through: $\delta_0 = e^\beta \sinh(\theta/2 + \vartheta/2)$ and $ \delta_1 = e^\beta \cosh(\theta/2 + \vartheta/2)$, the perturbed KG equation ~\eqref{eq:perts} is transformed into the dynamical system (see Appendix~\ref{pert_var}),

\begin{subequations}
\label{eq:deltas}
\begin{eqnarray}
\delta^\prime _0 &=&  \left[-3\sinh\theta-\frac{k^2}{k_J^2}(1 - \cosh \theta) \right]\delta_1-\frac{k^2}{k_J^2}\sinh \theta \delta_0 \nonumber \\ 
&& - \frac{\bar{h}'}{2}(1-\cosh\theta)\, , \label{d0p} \\
\delta^\prime_1 &=& \left(-3\cosh \theta+\frac{k^2_{eff}}{k_J^2}\sinh \theta  \right) \delta_1 - \frac{k^2_{eff}}{k_J^2}(1 + \cosh \theta) \delta_0 \nonumber \\ 
&& + \frac{\bar{h}'}{2} \sinh \theta \, . \label{d1p}
\end{eqnarray}
\end{subequations}

where $k_J^2 \equiv a^2 H^2 y_1$ is the (squared) Jeans wave number, and 
\begin{equation}
    k^2_{eff} \equiv a^2 H^2 \left( \frac{k^2}{a^2 H^2} + \frac{y_2}{2y} \Omega_\phi \right) \, . \label{eq:eqdeltas-c}
\end{equation}

In writing Eqs.~\eqref{eq:deltas} we have used the relation $\partial^2_\phi V = H^2 (y^2_1/4 - y y_2/2)$ in Eq.~\eqref{eq:perts}. Similarly to the case of scalar fields studied in~\cite{Cedeno:2017sou,LinaresCedeno:2020dte}, the variable $\delta_0$ is the phantom density contrast, as from Eqs.~\eqref{eq:hyper} and~\eqref{eq:linearvars} we find that $\delta \rho_\phi/\rho_\phi = (- \dot{\varphi} \dot{\phi} + \varphi \partial_\phi V)/\rho_\phi =\delta_0$.

Likewise, there is a Jeans wave number $k_J$ for the phantom density perturbations that only involves the function $y_1$~\cite{Urena-Lopez:2015odd,Cedeno:2017sou,LinaresCedeno:2020dte}. In the cases we will explore one expects that $y_1 \lesssim \mathcal{O}(1)$, and then the associated Jeans scale length will be equal or larger than the Hubble horizon, $k^{-1}_J \gtrsim 1/H$, which in general suggests that phantom perturbations will be suppressed in sub-horizon scales. It must be noticed that there is another scale involved in the evolution of the density perturbations, $k^2_{eff}$, which means that tachyonic effects will appear in phantom perturbations whenever $k^2_{eff} <0$~\cite{Cedeno:2017sou,LinaresCedeno:2020dte}, but this will depend on the chosen potential and the behavior of the combined variable $y_2 \Omega_\phi/y$. In general, phantom density perturbations are negligible, but we will include them in our study for completeness. 

\section{Phantom solutions \label{sec:phantom}}
The equations of motion~\eqref{eq:new4} can be closed if one writes down a functional form of the variable $y_2$. For purposes of simplicity, but also to ease the comparison with the quintessence case in~\cite{Urena-Lopez:2020npg,Roy:2018nce}, we take the following general parametrisation,
\begin{equation} 
y_2 = y \left( \alpha_0 + \alpha_1 y_1/y + \alpha_2 y^2_1/y^2 \right) \, . \label{eq:GP1}
\end{equation}

In doing so, we will be implicitly considering the same class of scalar potentials as in~\cite{Roy:2018nce} (see Tables~1 and~2 therein), as they are found from the functional relations of variables $y$, $y_1$ and $y_2$, which are independent of the nature of the field $\phi$. A similar parameterization of the phantom scalar field potentials has been suggested in~\cite{Roy:2017uvr}.

\subsection{Critical points}
To calculate the solutions of physical interest, in this section we start with the equations of the critical values $\theta_c$, $y_{1c}$ and $\Omega_{\phi c}$ as obtained from the dynamical system~\eqref{eq:new4}, namely 
\begin{subequations}
\label{eq:crit}
  \begin{eqnarray}
  -3 \sinh \theta_c - y_{1c} &=& 0 \label{eq:crita} \, , \\
  \frac{3}{2}\gamma_{tot} y_{1c} + \Omega_{\phi c}^{1/2} \sinh(\theta_c /2) y_{2c} &=& 0 \, , \label{eq:critb} \\
  3 (\gamma_{tot} - \gamma_{\phi c}) \Omega_{\phi c} &=& 0 \label{eq:critc} \, .
\end{eqnarray}
\end{subequations}

From Eq.~\eqref{eq:crita} we obtain the condition $y_{1c} = -3 \sinh \theta_c$, which is common to all possible critical points from Eqs.~\eqref{eq:crit}, and which will be also explicitly assumed in the analysis below for the phantom tracker and phantom dominated solutions in the following sections. Furthermore, if we consider Eq.~\eqref{eq:GP1}, then we get from Eq.~\eqref{eq:critb} either that $\sinh{\theta_{c}} =0$, or
\begin{equation}
    \gamma_{tot} - \frac{\alpha_0}{9} \Omega_{\phi c} + \frac{2}{3} \alpha_1 \Omega_{\phi c}^{1/2} \sinh(\theta_c/2) - 4 \alpha_2 \sinh^2{(\theta_c/2)} =0 \, . \label{eq:crit-sig}
\end{equation}

It is customary in the literature to classify the critical points that appear in the phantom equations of motion, in our case from Eqs.~\eqref{eq:crit} and~\eqref{eq:crit-sig}. The first critical point is the so-called fluid domination, for which $\Omega_{\phi c} =0$. One straightforward solution is $\sinh{\theta_{c}} =0$, which means that the phantom EoS takes the critical value $\gamma_{\phi c} = -1$. In contrast to the quintessence case, this time there is not kinetic dominated solution. Another possible solution under the condition $\Omega_{\phi c} =0$ is the tracker solution, but that is studied in more detail in Sec.~\ref{sec:phantom-tracker} below.

One final note is that there are not scaling solutions for phantom fields, in which the phantom EoS takes on the same values as that of the background dominant component $\gamma_\phi = \gamma_{tot}$, unless the background component is the cosmological constant or a phantom-like component too. 

\subsection{Phantom tracker solutions \label{sec:phantom-tracker}}
Let us first consider the case $\alpha_0 = 0= \alpha_1$, for which we obtain from Eq.~\eqref{eq:crit-sig} that the critical condition for the hyperbolic variable is $\sinh^2(\theta_{\phi,c}/2) = \gamma_{tot}/4\alpha_2$. In terms of the phantom EoS, the latter condition reads
\begin{equation}
 \gamma_{\phi,c} = -\gamma_{tot}/2\alpha_2 \, . \label{eq:tracker}
\end{equation}

Notice that in Eq.~\eqref{eq:tracker} we must choose positive definite values for $\alpha_2$ so that $\gamma_{\phi,c} < 0$. Moreover, a quick comparison with previous studies confirms that Eq.~\eqref{eq:tracker} is the tracker condition for phantom fields. 

The potentials that exhibit the tracker behavior according to Eq.~\eqref{eq:tracker} are of the power-law form $V(\phi) = M^{4-p} \phi^p$, where $p = 2/(1+2\alpha_2)$. In contrast to the quintessence in which the tracker potentials are of the inverse-power law type, this time the tracker condition is achieved for $0 < p < 2$ (corresponding to $0 < \alpha_2 < \infty$), which means that the phantom field evolves away from the minimum of the potential while in the tracker regime.

As argued in~\cite{Urena-Lopez:2020npg}, the tracker condition~\eqref{eq:tracker} is of wider applicability if $(\alpha_0, \, \alpha_1) \neq 0$, as long as $\Omega_{\phi c}$ is negligible, which is generically expected at early times. Moreover, if we can write $y_2 = y f(y_1/y)$, where $f$ is an arbitrary function of its argument, then the critical condition~\eqref{eq:crit-sig} reads
\begin{equation}
   \left[ 9 \gamma_{tot} + \Omega_{\phi c} \, f \left( \frac{3 \sqrt{2} \sinh(\theta_c/2)}{\Omega^{1/2}_{\phi c}} \right) \right] \sin \theta_c =0 \, . \label{eq:crit-gen}
\end{equation}
In writing Eq.~\eqref{eq:crit-gen} we have used $y_{1c}/y_c = 3 \sinh \theta_c/[\Omega^{1/2}_{\phi c} \cosh(\theta_c/2)] = 3 \sqrt{2} \sinh(\theta_c/2)/\Omega^{1/2}_{\phi c}$. Thus, the tracker solution exists whenever the following condition is satisfied,
\begin{equation}
    \lim_{\Omega_{\phi c} \to 0} \left[ \Omega_{\phi c} \, f \left( \frac{3 \sqrt{2} \sinh(\theta_c/2)}{\Omega^{1/2}_{\phi c}} \right) \right] = g(\sinh(\theta_c/2)) \, , \label{eq:general-tracker}
\end{equation}
where $g(x)$ would be the resultant function after the limit operation. The tracker equation derived from Eq.~\eqref{eq:crit-gen} under the result~\eqref{eq:general-tracker} would simply read: $9 \gamma_{tot} + g (\sinh(\theta_c/2)) = 0$. Any valid solution of the latter equation should be considered a generalized tracker solution for the phantom field.

\subsection{Phantom dominated solutions \label{sec:phantom-dominated}}
Let us turn our attention to phantom dominated solutions at late times; these solutions are characterised by the conditions $\Omega_{\phi c} = 1$ and $\gamma_{tot} = \gamma_{\phi c}$. Our main interest here are the phantom dominated solutions that are related to the tracker solutions at early times.

A small note is in turn. The phantom EoS, given by $\gamma_\phi = -2\sinh^2 (\theta_c/2)$, is the same irrespective of the sign of $\theta$, but because $\gamma_\phi \leq 0$, $\theta$ does not cross the zero value and then one needs to choose either the negative or positive branch of the hyperbolic sine. For convenience, we will hereafter choose the negative branch, $\theta  \leq 0$, which also allows for the potential variable $y_1$ to be positive definite.

Recalling that the first option for a critical value is $\sinh \theta_c = 0$, we find that one possible asymptotic value of the phantom EoS is $\gamma_{\phi c}=0$, for which the phantom density is dominated by its potential part $V(\phi)$. This means that at late times the phantom field approaches the behavior of a cosmological constant.

Another possibility arises from the solution of Eq.~\eqref{eq:crit-sig}, which for the aforementioned conditions of phantom domination reads
\begin{equation}
     \alpha_0 - 6 \alpha_1 \sinh (\theta_c/2) \,  + 18 (1 + 2 \alpha_2) \sinh^2 (\theta_c/2) =0 \, . \label{eq:phantom-dom}
\end{equation}
The critical solutions of Eq.~\eqref{eq:phantom-dom} will depend on the values of the active parameters $\alpha$. For the particular case of purely tracker solutions, $\alpha_0 = 0 = \alpha_1$ the only critical solution possible is again $\theta_c =0$, and then $\gamma_{\phi c} = 0$, which means that the phantom density will asymptotically behave as a cosmological constant.

We now study the conditions for Eq.~\eqref{eq:phantom-dom} to have at least one negative solution, that is, $\theta_c <0$, under the tracker condition $\alpha_2 >0$. Let us start with $\alpha_1 =0$, for which the solution of Eq.~\eqref{eq:phantom-dom} is
\begin{subequations}
\label{eq:sol-pdom}
\begin{equation}
\sinh (\theta_c/2) = \pm \left[ - \frac{\alpha_0}{18(1 + 2 \alpha_2)} \right]^{1/2} \, . \label{eq:sol-pdoma}
\end{equation}
It is clear that $\alpha_0 <0$ is required to have a negative real solution and then also $\gamma_{\phi c} <0$. 

In the case $\alpha_1 \neq 0$, the general solution of Eq.~\eqref{eq:phantom-dom} can be written in the form
\begin{equation}
    \sinh (\theta_c/2) = \frac{\alpha_1 \pm |\alpha_1| \sqrt{\Delta}}{6(1+2\alpha_2)} \, , \label{eq:sol-pdomb}
\end{equation}
\end{subequations}
where $\Delta = 1 - 2 \alpha_0 (1+2\alpha_2)/\alpha^2_1$. Notice that we require $\Delta \geq 0$ to have real valued solutions of Eq.~\eqref{eq:sol-pdomb}.

We first consider the case $\alpha_0 < 0$. The latter implies that $\Delta > 1$, which then assures the existence of at least one negative solution of Eq.~\eqref{eq:sol-pdomb}, irrespective of the value of $\alpha_1$. In other words, a negative value of $\alpha_0$ assures the existence of a phantom EoS at late times for tracker potentials. Next, we take the case $\alpha_0 \geq 0$, such that $0 \leq \Delta \leq 1$. There will be at least one negative solution of Eq.~\eqref{eq:sol-pdomb}, and then again a phantom EoS, if $\alpha_1 < 0$.

In summary, one consequence of our choice $\theta  \leq 0$ is that the cosmological constant case ($\gamma_\phi =0$) is the only asymptotic solution available for the phantom EoS if both conditions $\alpha_0 \geq 0$ and $\alpha_1 \geq 0$ are satisfied, as for such conditions there are not negative solutions of $\theta_c$ from Eq.~\eqref{eq:phantom-dom}. In all other cases, as long as $\Delta \geq 0$, the phantom EoS remains below the phantom divide ($\gamma_\phi < 0$) and its asymptotic value is given by the negative solution of Eqs.~\eqref{eq:sol-pdom}.

\section{\label{sec:tracker}Numerical solutions}
To obtain reliable numerical solutions of the phantom equations of motion, we use the tracking condition~\eqref{eq:tracker} discussed above to find a set of initial conditions that can be related to the current observed values of the cosmological parameters. The resultant equations are 
\begin{subequations}
\label{eq:initial}
\begin{eqnarray}
\cosh \theta_i &=& 1 + \frac{2}{3\alpha_2} \, , \quad y_{1i} = -3 \sinh \theta_i \, , \label{eq:initial-a} \\ 
\Omega_{\phi i} &=& A \times a^{4(1+1/2\alpha_2)}_i \left( \frac{\Omega_{m 0}}{\Omega_{r 0}} \right)^{1+1/2\alpha_2} \Omega_{\phi 0} \, , \label{eq:initial-b}
\end{eqnarray}
\end{subequations}
where $\Omega_{r0}$, $\Omega_{m0}$ and $\Omega_{\phi 0}$ are, respectively, the present density parameters of relativistic matter, nonrelativistic matter and the phantom field. The initial value of the scale factor is given by $a_i$ which typically considered to be  $a_i \simeq 10^{-14}$. The initial conditions for the variables $\theta$ and $y_1$ are obtained directly from the tracking condition~\eqref{eq:tracker}, whereas the initial value of $\Omega_{\phi i}$ is obtained from the integration of the background equation~\eqref{eq:new4b} for the radiation and matter domination epochs.

For the numerical solutions, we rely on an amended version of the the Boltzmann code \textsc{class} (v2.9)~\cite{Lesgourgues:2011rg,*Blas:2011rf,*Lesgourgues:2011re,*Lesgourgues:2011rh}, which internally adjusts the value of the constant coefficient $A$, so that the desired value of the phantom density parameter $\Omega_{\phi 0}$ at present is obtained. For the initial conditions of the linear perturbations, we simply use $\delta_{0i} = 0$ and $\delta_{1i} = 0$, as the evolution of the perturbation variables is mostly driven by the nonhomogeneous terms in Eqs.~\eqref{eq:deltas}.

\subsection{Phantom dark energy ($\phi$)}
\begin{figure}[htp!]
\includegraphics[width=0.49\textwidth]{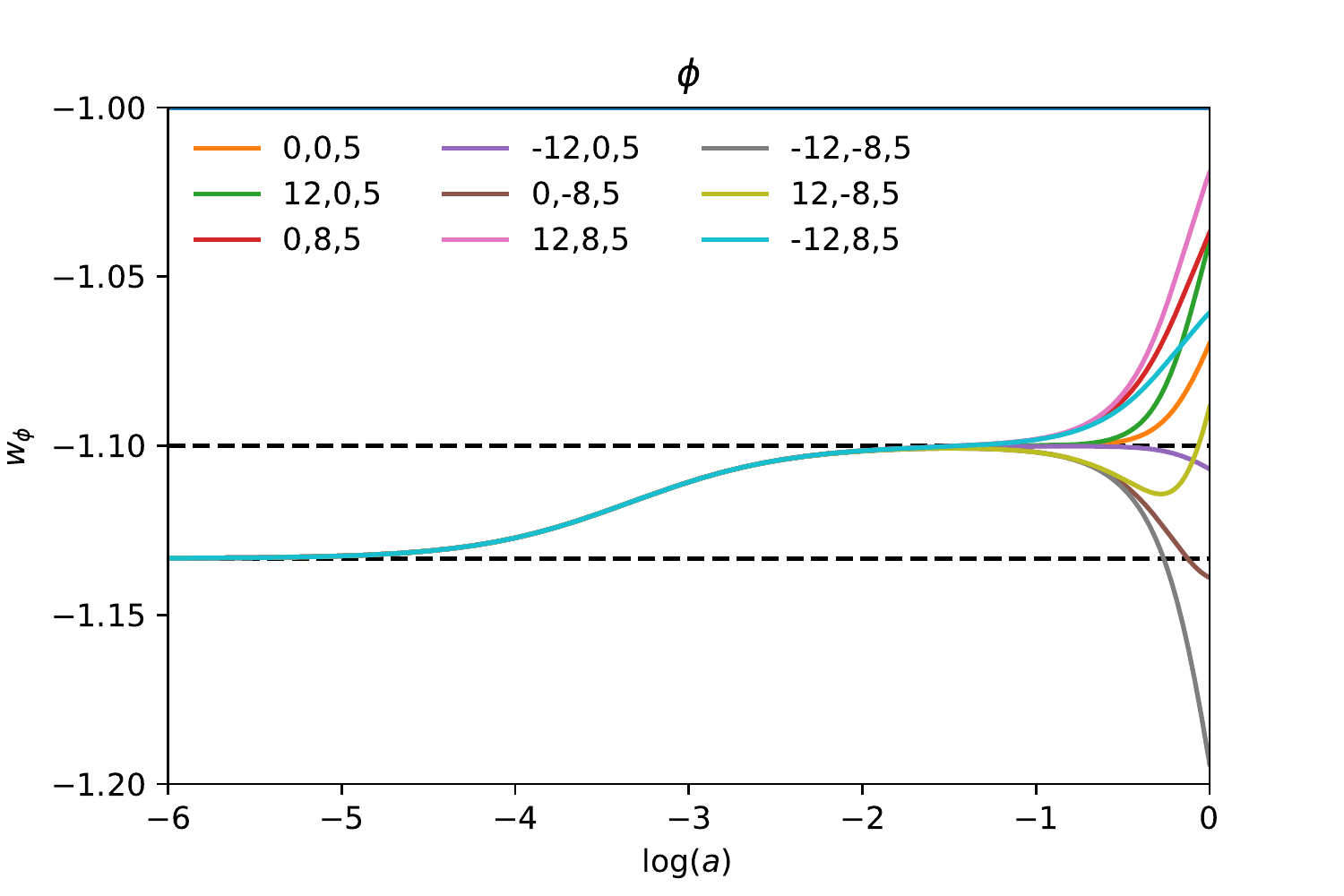}
\includegraphics[width=0.49\textwidth]{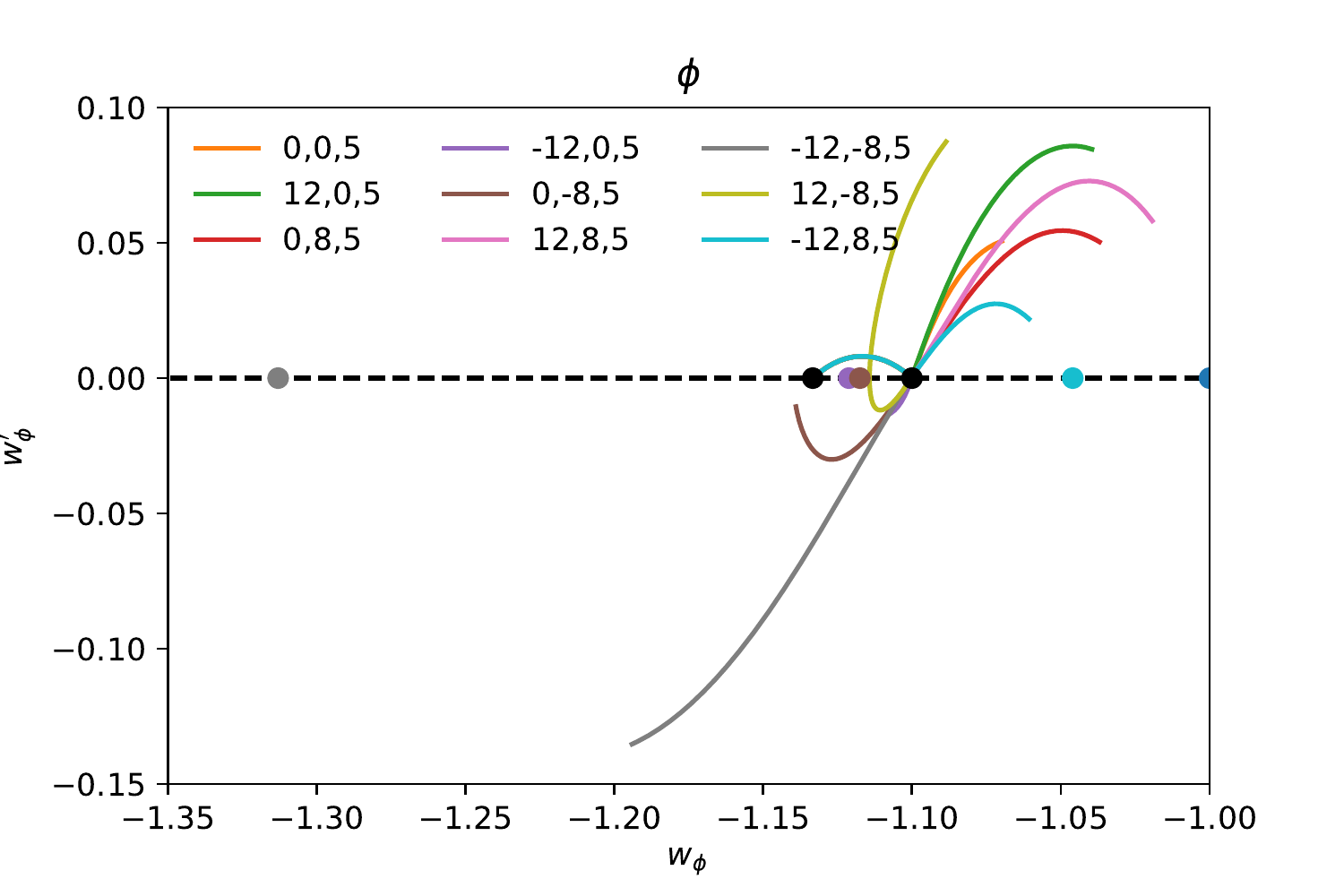}
\caption{\label{fig:numerics-a} (Top) The evolution of the EoS $w_\phi$ for tracker phantom models, with the values of the active parameters as indicated by the triplets $(\alpha_0,\alpha_1,\alpha_2)$ in the labels of the curves (for the corresponding potentials see Table~II in ~\cite{Roy:2018nce}). (Bottom) Phase space behaviour of the phantom EoS, in the plane $(w_\phi,w^\prime_\phi)$, for the same cases (with the same colors) as in the top panel. The blue dot corresponds to the cosmological constant case, whereas the black dots represent the tracker values at $(-17/15,0)$ and $(-11/10,0)$, corresponding to the dashed black lines in the top panel. The grey, purple, brown and light-blue dots indicate the solutions of Eq.~\eqref{eq:phantom-dom}, which are also the asymptotic points for the corresponding curves of the same color. See the text for more details.}
\end{figure}

Here we study purely phantom solutions, and then $\rho_\Lambda =0$; we label this case as $\phi$. Typical examples for the behavior of the phantom EoS are shown in Fig.~\ref{fig:numerics-a} for the fixed value $\alpha_2=5$, together with different combinations of the other active parameters $\alpha_0$ and $\alpha_1$. Other relevant parameters, like the present density contributions of the different matter species, were fixed to the values reported by the Planck collaboration (see their Table~1)~\cite{Aghanim:2018eyx}.

In the top panel of Fig.~\ref{fig:numerics-a}, it can be seen that all solutions maintain their tracker behavior at early times, as seen from the values of the phantom EoS during the radiation and matter domination epochs, which are  $-17/15$ and $-11/10$ (dashed black lines), respectively. The evolution of the solutions from radiation to matter domination for the different examples are so identical that they are not distinguishable in the plot. Recalling that the initial conditions are set up at $a_i = 10^{-14}$, this indicates that the tracker condition~\eqref{eq:tracker} is a stable solution of the background evolution at early times.

As for late times, a better view of the evolution of the phantom EoS is provided in the bottom panel of Fig.~\ref{fig:numerics-a}, in terms of the phase space $(w_\phi,w^\prime_\phi)$, where $w^\prime_\phi = \sinh \theta (3 \sinh\theta + y_1)$. All solutions depart from the tracker point at radiation domination $(-17/15,0)$ (left black dot), and evolve towards that at matter domination $(-11/10,0)$ (right black dot), while tracking the background and with identical evolutionary paths. Again, for the cases in which $\alpha_0 > 0$ and $\alpha_1 > 0$, see Sec.~\ref{sec:phantom-dominated} above, the curves are deflected away from the second tracker point and the phantom EoS evolves towards the cosmological constant point at $(-1,0)$. 

For all other cases, the asymptotic values of the phantom EoS are also indicated by dots in the bottom panel of Fig.~\ref{fig:numerics-a}, with the same color as that of the corresponding evolution curve. The coordinates of the asymptotic points (brown, purple, grey, and light blue dots) were obtained from the solutions of Eqs.~\eqref{eq:phantom-dom}.

\begin{figure}[htp!]
\includegraphics[width=0.49\textwidth]{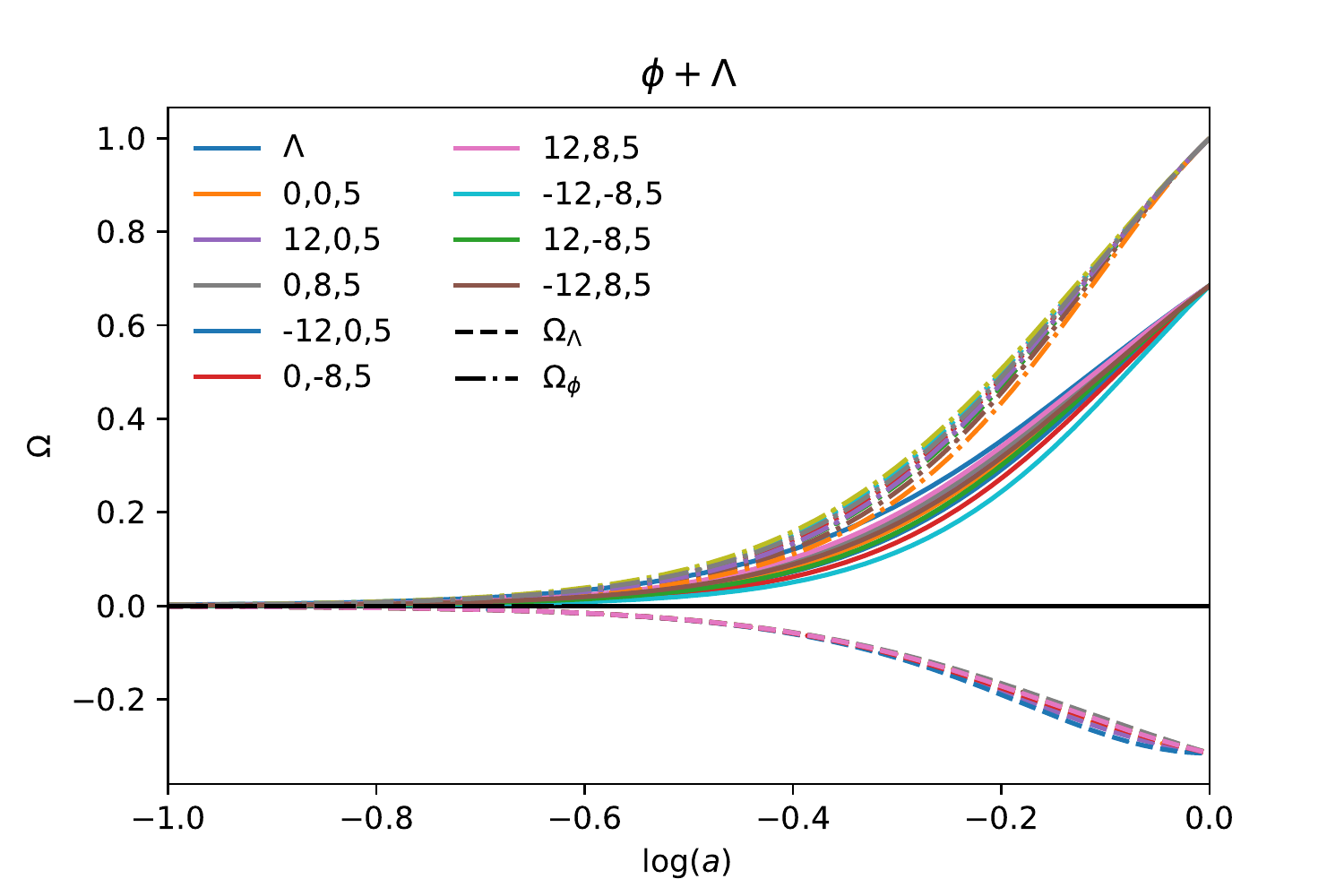}
\includegraphics[width=0.49\textwidth]{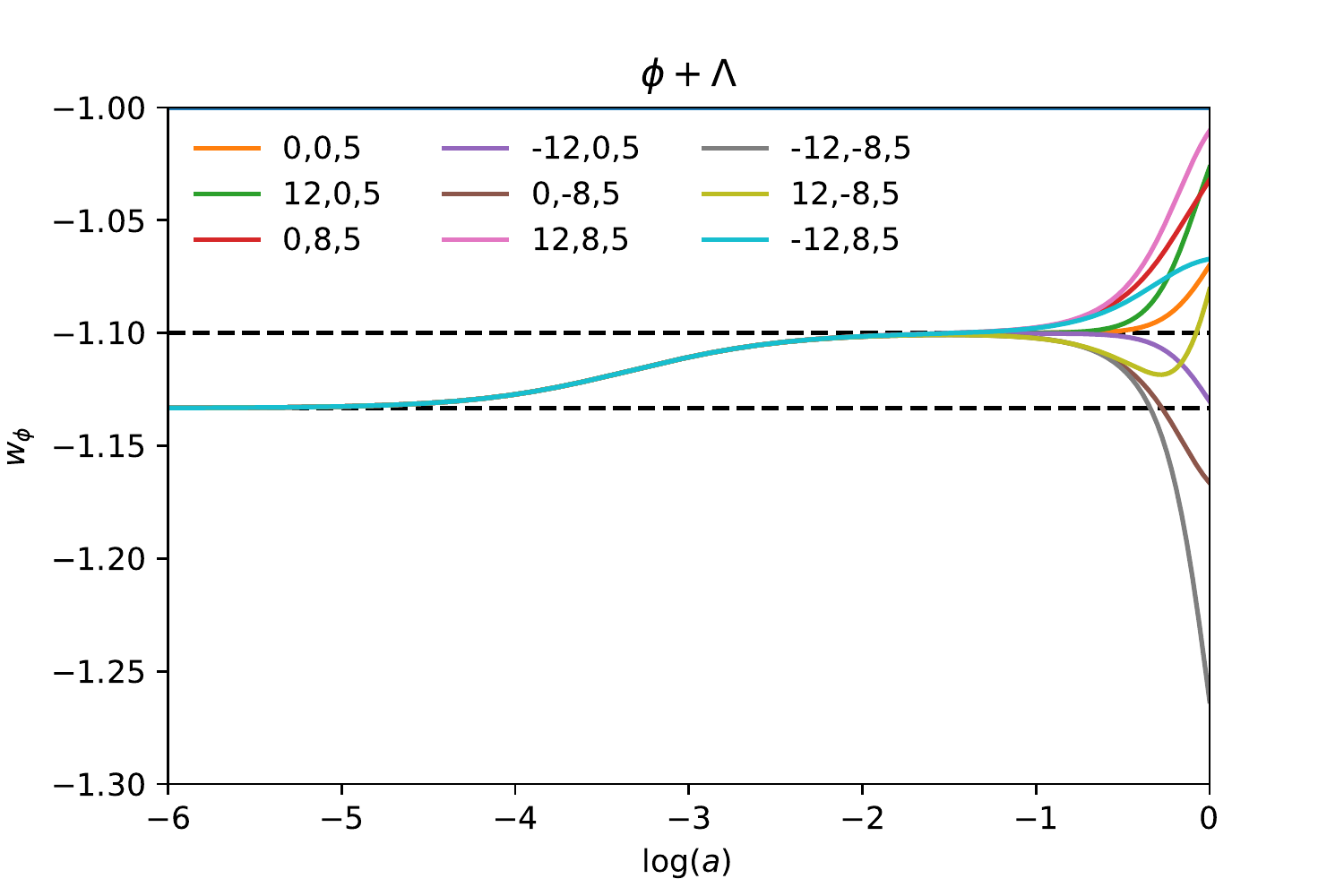}
\includegraphics[width=0.49\textwidth]{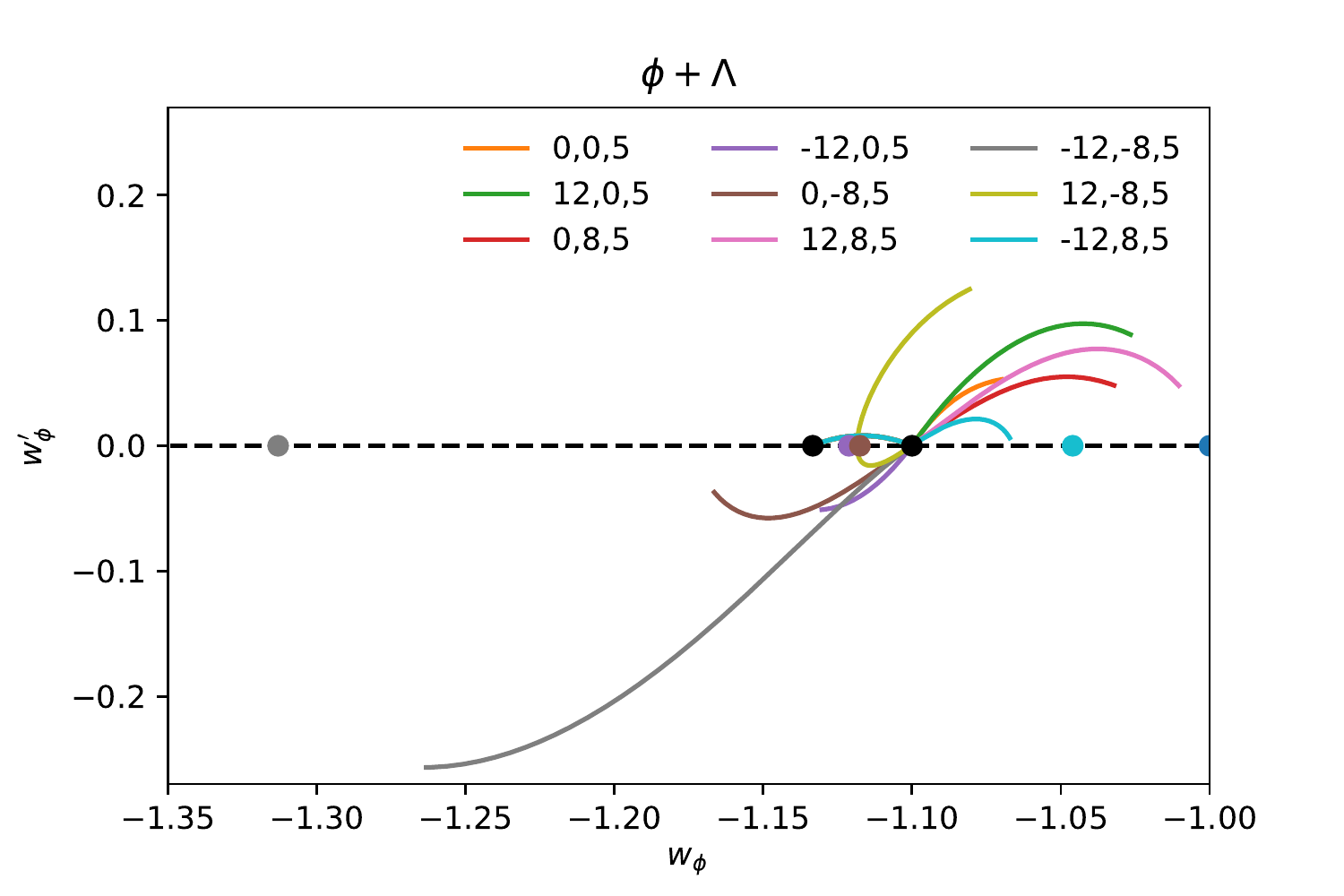}
\caption{\label{fig:numerics-c} (Top panel) Late-time evolution of the density parameters of the phantom field $\Omega_\phi$ (dot-dashed curves), the cosmological constant $\Omega_\Lambda$ (dashed curves) and the total DE budget (solid curves). (Middle panel) The evolution of the EoS $w_\phi$ for the same cases as in the top panel. (Bottom panel) Phase space behaviour of the EoS $(w_\phi,w^\prime_\phi)$ for the same cases  as in the top  and the middle panels. The dots in the bottom panel have the same meaning as in the bottom panel of Fig.~\ref{fig:numerics-a} above. See the text for more details.}
\end{figure}

\begin{figure*}[htp!]
\includegraphics[width=0.49\textwidth]{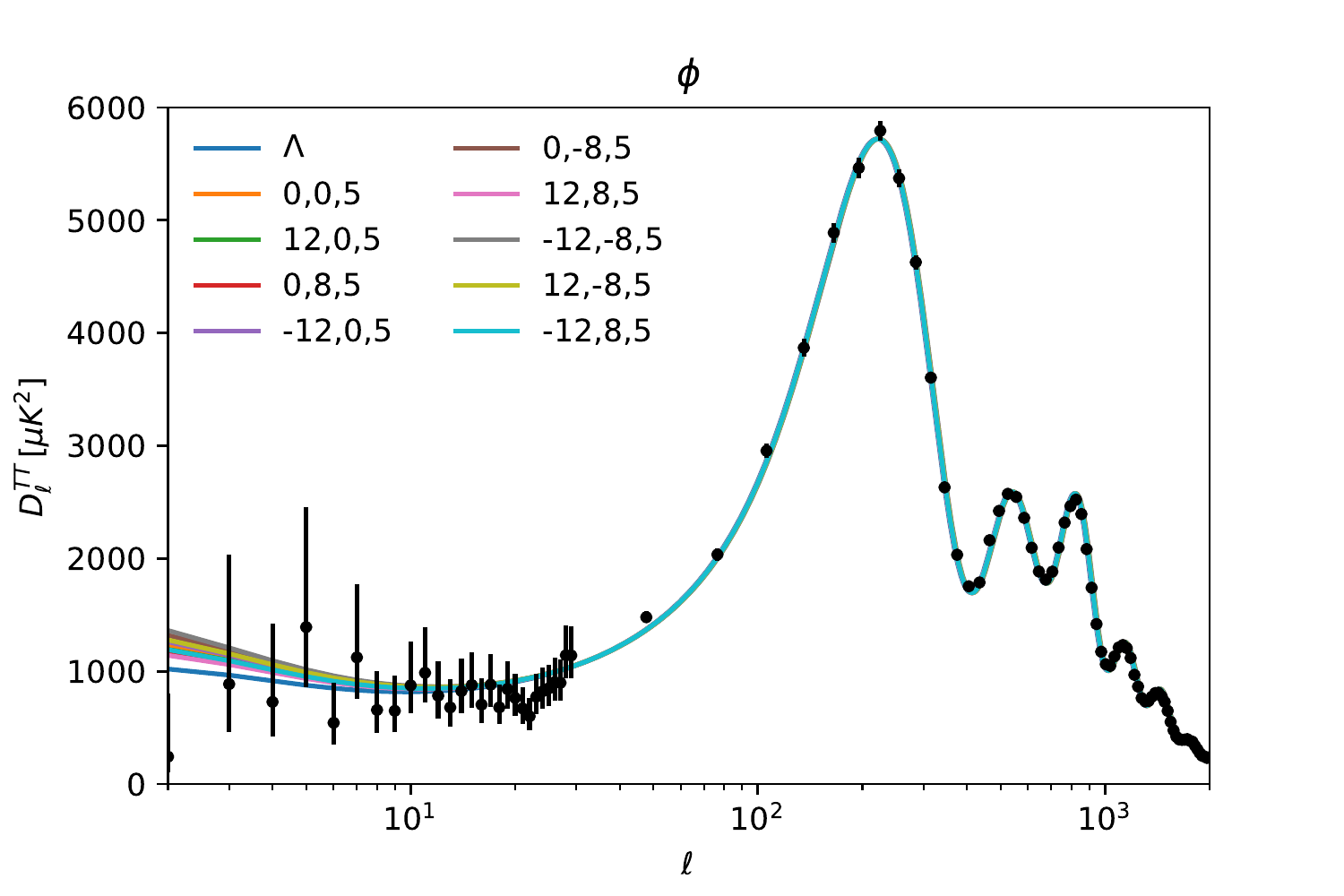}
\includegraphics[width=0.49\textwidth]{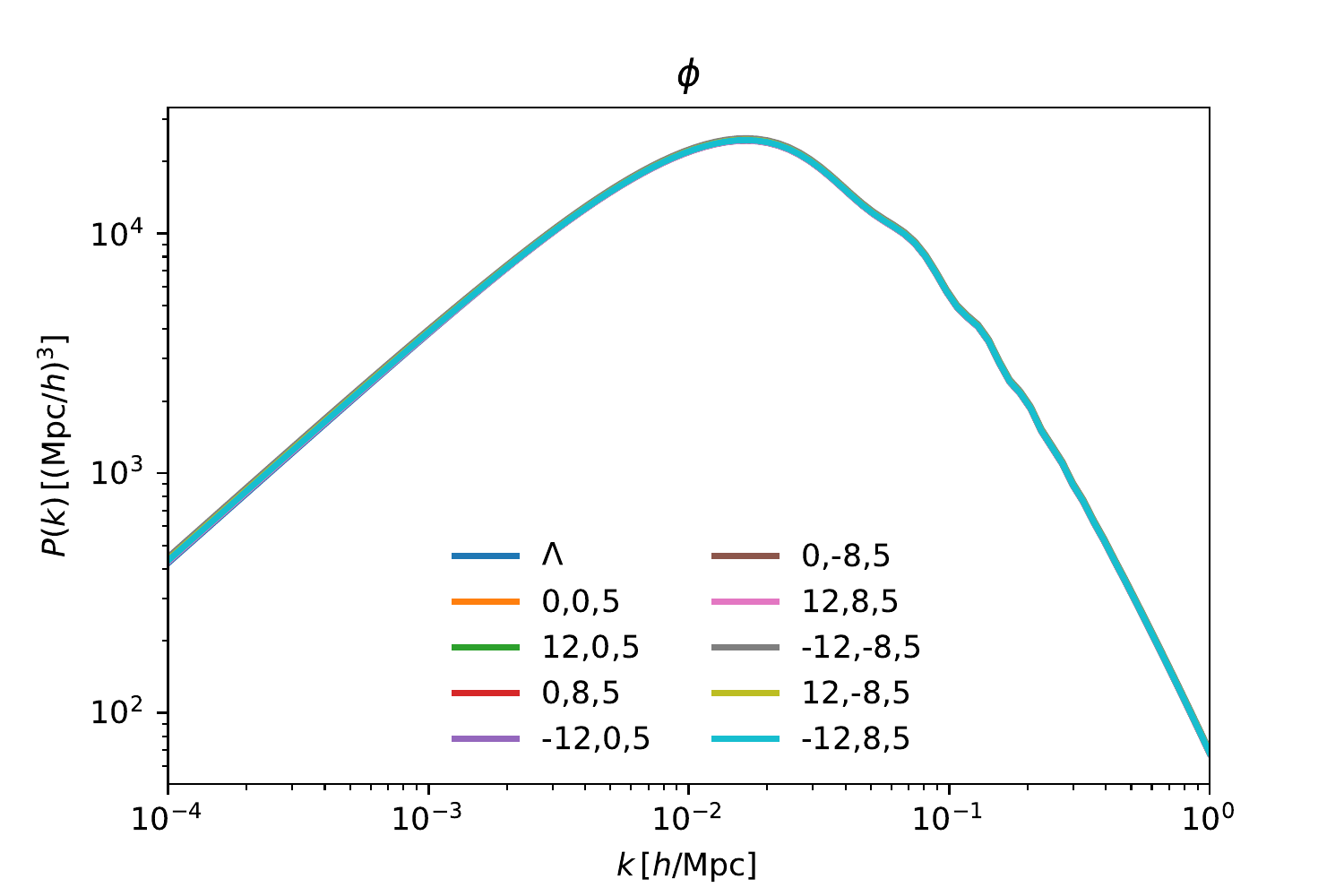}
\includegraphics[width=0.49\textwidth]{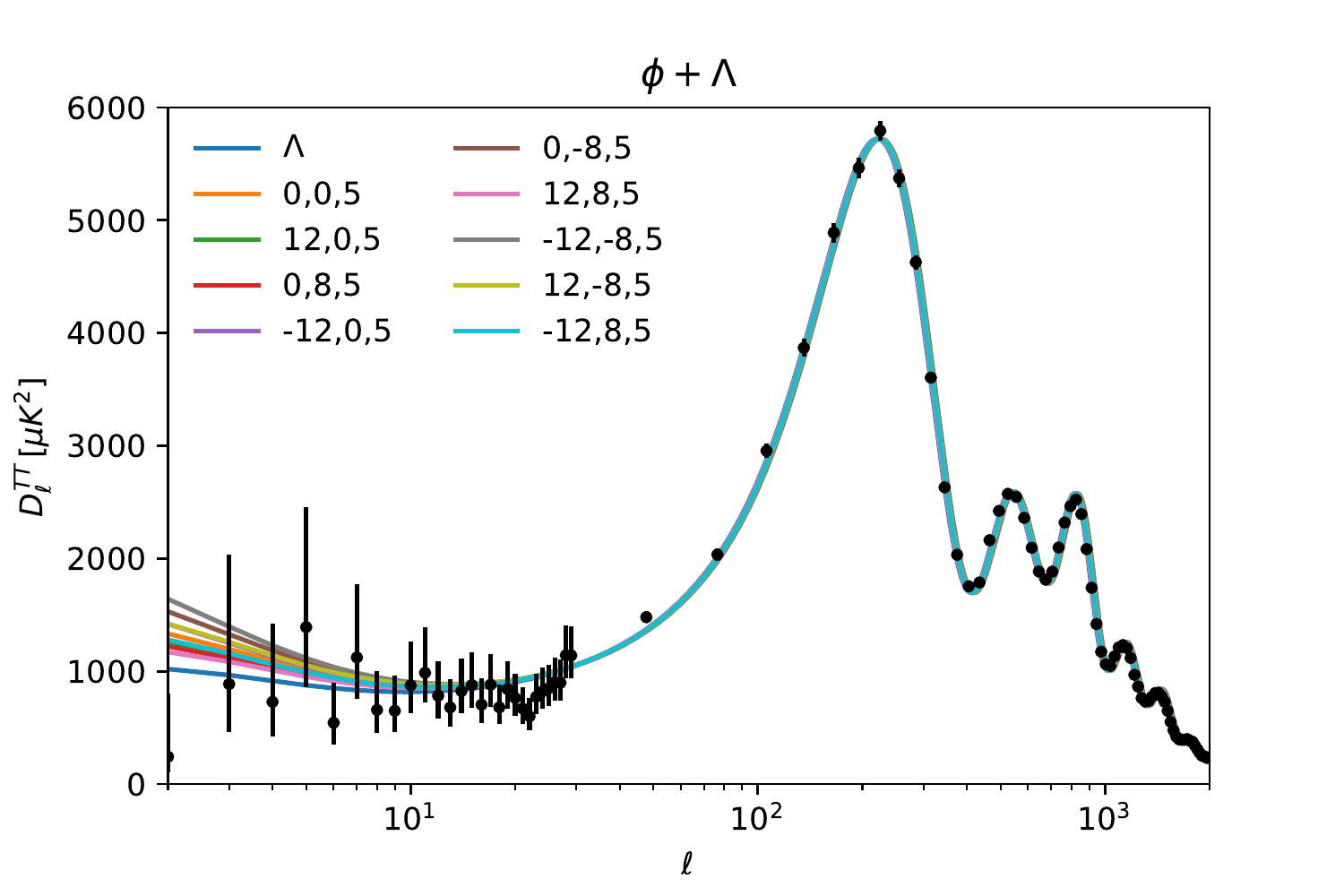}
\includegraphics[width=0.49\textwidth]{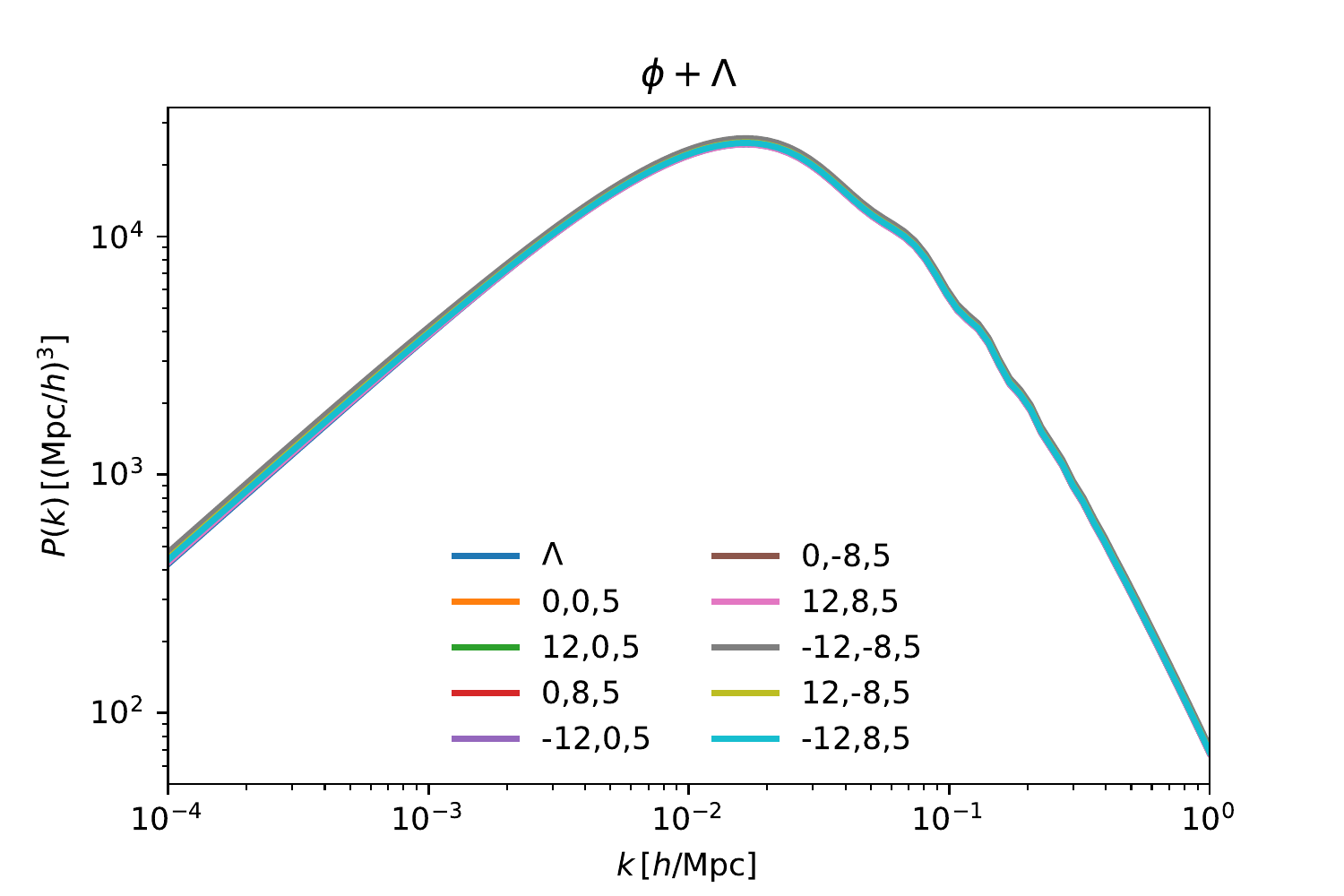}
\caption{\label{fig:numerics-b} The anisotropies of the CMB  and the MPS for the same models shown in Fig.~\ref{fig:numerics-a} (top panels, Phantom DE) and in Fig~\ref{fig:numerics-c} (bottom panels, Phantom-$\Lambda$ DE). The case of $\Lambda$CDM is also shown for reference in each case. The dots in the top panel are the binned TT power spectrum from the Planck collaboration. See the text for more details.}
\end{figure*}

To show the influence of phantom density perturbations in models of phantom DE, we show in the top panels of Fig.~\ref{fig:numerics-b} the two-point temperature power spectrum $\mathcal{C}^{TT}_\ell$ of the cosmic microwave background (CMB) and the mass power spectrum (MPS) of linear density perturbations $P(k)$, for the same numerical examples shown in Fig.~\ref{fig:numerics-a}. In comparison with the standard case with $\Lambda$ as DE, we see that there are noticeable changes, specially for the CMB spectrum, but only at large scales and for the most extreme phantom values of the DE EoS.

\subsection{Phantom-$\Lambda$ dark energy ($\phi+\Lambda$)}
We now turn our attention to the case in which both the cosmological constant and the phantom field are part of the DE budget, a case we label as $\phi+\Lambda$. For a comparison with the phantom case in the previous section, we show in Fig.~\ref{fig:numerics-c} the evolution of the density parameters $\Omega_\phi$ and $\Omega_\Lambda$ at recent times, together with the phase space of the phantom EoS for the same triplets (and colors) $(\alpha_0,\alpha_1,\alpha_2)$ as in Fig.~\ref{fig:numerics-a}. 

For all plots, we chose $\Omega_\phi = 1.0$. As the present density contributions of the different matter species are fixed to the values reported by the Planck collaboration, the present value of $\Omega_{\Lambda}$ was adjusted so as to fulfill the Friedmann constraint for a flat Universe. For this reason the contribution of the cosmological constant is in general negative, see the top panel of Fig.~\ref{fig:numerics-c}. 

Notice that the corresponding behavior of the phantom EoS, as shown in the middle and bottom panels of Fig.~\ref{fig:numerics-c}, is qualitatively the same as in the standard phantom case in Fig.~\ref{fig:numerics-a}, the only difference being that the present EoS seems to reach more negative values than in the phantom-only case.

As for the phantom perturbations, we also show in the bottom row of Fig.~\ref{fig:numerics-b} the temperature anisotropies and the MPS for the same cases shown in Fig.~\ref{fig:numerics-c}. There is an enhancement of the power at large scales in the two observables, which seems to be an effect of the larger contribution of the phantom field to the DE budget, and also of the respective smaller influence of the (negative) cosmological constant.

\section{\label{sec:numerical}Comparison with observations}
Here we present the constraints on the phantom tracker models arising from the comparison with cosmological observations. For this, we used the aforementioned Boltzmann code \textsc{class} and the MCMC sampler \textsc{monte python} (v3.3), together with the following observations: Pantheon, BAO (BOSS DR12~\cite{Alam_2017}, 6dFGS~\cite{Beutler_2011}, eBOSS DR14 (Lya)~\cite{Cuceu_2019}, and WiggleZ~\cite{Kazin:2014qga}), SH0ES and a compressed Planck likelihood. For completeness, we also included observations about cluster counts (SDSS LRG DR7~\cite{Ross_2015}, SDSS LRG DR4~\cite{Tegmark_2006} and WiggleZ~\cite{Kazin:2014qga}) to put constraints on possible changes on the MPS because of the phantom density perturbations (see Fig.~\ref{fig:numerics-c} above).

For the compressed Planck likelihood, we considered the proposal in~\cite{Arendse_2020} (see their Appendix~A) for the baryon physical density $\omega_b = \Omega_b h^2$ and the two shift parameters,
\begin{equation}
    \theta_\ast = r_s(z_{dec})/D_A(z_{dec}) \, , \quad \mathcal{R} = \sqrt{\Omega_M H^2_0} D_A(z_{dec}) \, ,
\end{equation}
where $z_{dec}$ is the redshift at decoupling and $D_A$ is the comoving angular diameter distance. As stated in~\cite{Arendse_2020}, we have also verified that we recover the standard Planck constraints on a flat $\Lambda$CDM model from the compressed likelihood.

The sampled parameters and their corresponding flat priors were as follows. For the physical baryon density, $100 \omega_b = [1.9,2.5]$, for the physical CDM density $\omega_{cdm} = [0.095,0.145]$, and the Hubble parameter $H_0=[60,74] \, \mathrm{km \, s^{-1} \, Mpc^{-1}}$. Following the standard prescription in \textsc{class}, the present contributions of the DE components are determined the last from the closure of the Friedmann constraint for a flat Universe. In particular for the Phantom+$\Lambda$ case, and for numerical convenience, we sampled the phantom parameter in the range $\Omega_{\phi} = [0.1,1]$, and the contribution from $\Lambda$ was calculated from the Friedmann constraint~\eqref{eq:fri_a}. 

Finally, the phantom free parameters were sampled in the ranges $\alpha_0 = [-12,12]$, $\alpha_1 = [-8,8]$ and $\alpha_2 = [1,16]$. These ranges were chosen to ease the shooting procedure that determines the present value of $\Omega_{\phi}$, but are also in agreement with the expected values on typical potentials in the literature. See, for instance Table~1 in Ref.~\cite{Roy:2018nce}, where the active parameters of the listed potentials are all of the order of unity.

\subsection{General constraints on model parameters}
\begin{figure}[htp!]
\includegraphics[width=0.49\textwidth]{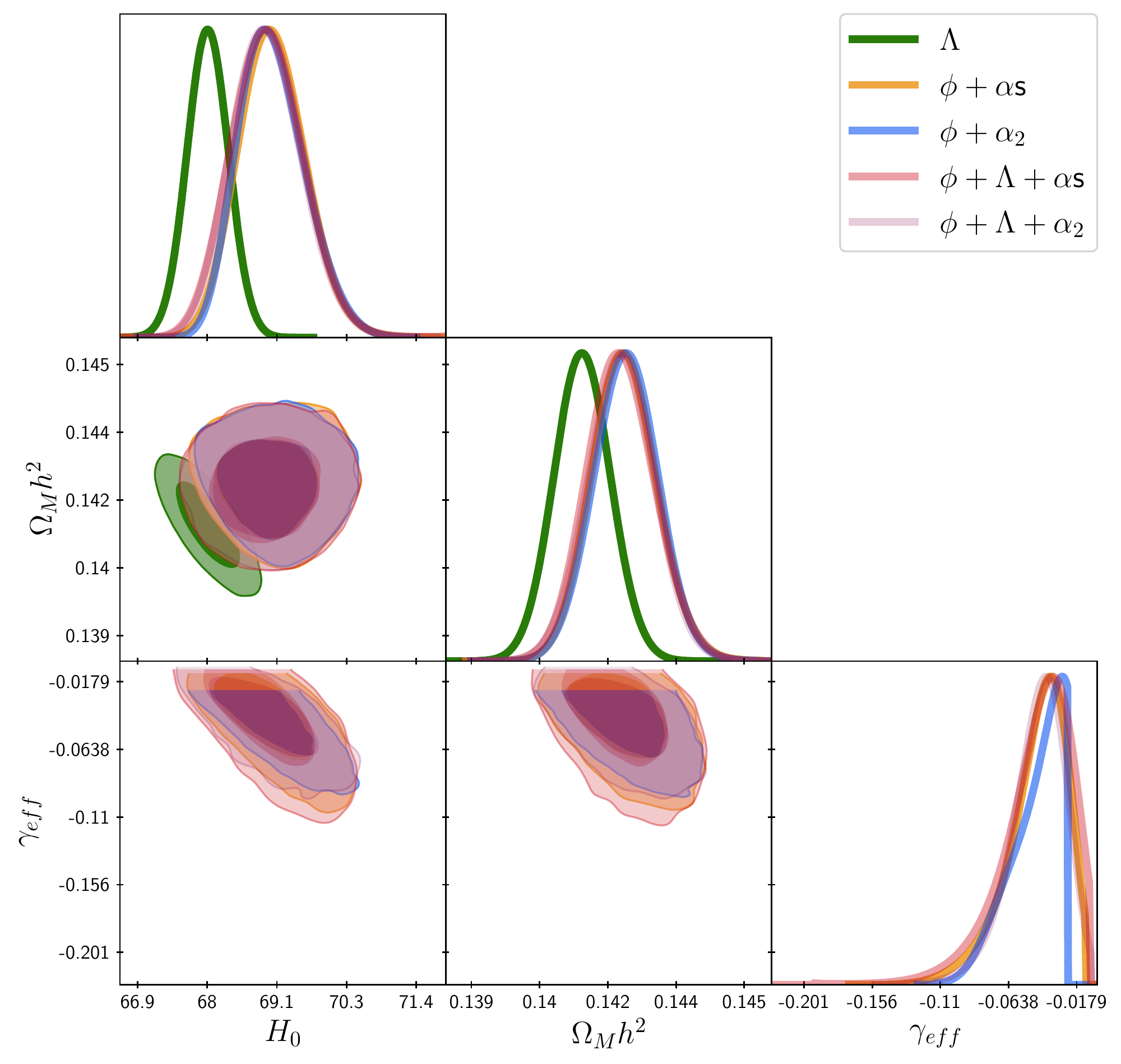}
\caption{\label{fig:general} Observational constraints on $H_0$, $\Omega_M h^2$ and $\gamma_{eff}$, for the same type of tracker potentials shown in Figs.~\ref{fig:numerics-a}, \ref{fig:numerics-c} and~\ref{fig:numerics-b}. The labels in the plots correspond to the models in Table~\ref{tab:params}. See the text for more details.}
\end{figure}

The obtained constraints on the models are shown in Fig.~\ref{fig:general}, with their detailed values listed in Table~\ref{tab:params}. Also, the models considered were labeled as: $\Lambda$ (the cosmological constant), $\phi$ (phantom DE) and $\phi+\Lambda$ (phantom and a cosmological constant). For the latter two, we have two further sub-cases: the purely tracker solution labeled as $\phi + \alpha_2$, and the generalized one $\phi + \alpha$'s.

In Fig.~\ref{fig:general} we show the confidence regions for the Hubble parameter $H_0$, the physical matter density $\Omega_M h^2$, and the effective DE equation of state $\gamma_{eff}$ at the present time. We first note that the obtained values of $H_0$ and $\Omega_M h^2$ are the same for the phantom models, which is expected from the strong constraints imposed by the compressed Planck likelihood on these parameters, even in the presence of other late-time observations. 

\begin{table*}[htp!]
\caption{\label{tab:params} Fitted values of the different models described in the text. The confidence regions for the parameters are shown in Figs.~\ref{fig:general} and~\ref{fig:alphas}, using the same labels for each case. The last column is for the model $\phi + \Lambda +\alpha_2$ with the extended prior $\Omega_\phi = [0.1:2]$. $k$ is the number of extra parameters with respect to $\Lambda$ case. See the text for more details.}
\begin{ruledtabular}
\begin{tabular}{|c|c|c|c|c|c|c|}
Parameter & $\Lambda$ & $\phi + \alpha$'s & $\phi+\alpha_2$ & $\phi + \Lambda +\alpha$'s & $\phi + \Lambda +\alpha_2$ & $\phi + \Lambda +\alpha_2$ (Ext.) \\ 
\hline
$H_0$ & $68.0_{-0.3}^{+0.3}$ & $69.1_{-0.6}^{+0.5}$ & $69.1_{-0.6}^{+0.5}$ & $69.0_{-0.6}^{+0.6}$ & $69.0_{-0.6}^{+0.5}$ & $69.28_{-0.62}^{+0.63}$ \\
\hline
$\Omega_M h^2$ & $0.141_{-0.0007}^{+0.0007}$ & $0.142_{-0.0009}^{+0.0008}$ & $0.142_{-0.0008}^{+0.0008}$ & $0.142_{-0.0009}^{+0.0008}$ & $0.142_{-0.0008}^{+0.0008}$ & $0.1426_{-0.0009}^{+0.00088}$ \\
\hline
$\gamma_{eff}$ & $0$ & $-0.045_{-0.012}^{+0.026}$ & $-0.045_{-0.006}^{+0.024}$ & $-0.045_{-0.014}^{+0.030}$ & $-0.044_{-0.014}^{+0.022}$ & $-0.04792_{-0.014}^{+0.017}$ \\
\hline
$\Omega_\Lambda$ & $0.694_{-0.0043}^{+0.0046}$ & $0$ & $0$ & $0.0462_{-0.317}^{+0.144}$ & $0.0371_{-0.315}^{+0.133}$ & $-0.3504_{-0.4}^{+0.56}$ \\
\hline
$\Omega_\phi$ & $0$ & $0.7013_{-0.0047}^{+0.0048}$ & $0.7012_{-0.0051}^{+0.0046}$ & $0.6249_{-0.12}^{+0.37}$ & $0.9138_{-0.56}^{+0.34}$ & $1.053_{-0.56}^{+0.4}$ \\
\hline
$\alpha_2$ & $0$ & $8.99_{-4.7}^{+3.0}$ & $8.78_{-4.68}^{+2.02}$ & $8.47_{-0.12}^{+0.37}$ & $8.56_{-4.51}^{+3.65}$ & $10.48_{-1.7}^{+5.5}$ \\
\hline
$k$ & $0$ & $+3$ & $+1$ & $+4$ & $+2$ & $+1$ \\
\hline
$\Delta \chi^2_{min}$ & $0$ & $-6$ & $-5$ & $-5$ & $-5$ & $-4$ \\
\hline
$\ln B_{\phi \Lambda}$ & $0$ & $+2.51$ & $+2.13$ & $+2.27$ & $+2.05$ &  $+2.05$ \\
                    & & Definite/Positive & Definite/Positive & Definite/Positive & Definite/Positive & Definite/Positive
\end{tabular}
\end{ruledtabular}
\end{table*}

Note that there is a noticeable shift in the central values of both parameters as compared to the case of $\Lambda$, which is an effect that only appears when late-time observations are included in the analysis. However, the shift in the Hubble parameter in the phantom models is far from solving the Hubble tension with the SH0ES measurement.

The effective barotropic EoS of the whole DE budget is explicitly defined as:
\begin{equation}
    \gamma_{eff} \equiv \frac{1}{\sum_j \Omega_j} \sum_j \gamma_j \Omega_j \, ,
\end{equation}
where the index $j$ only runs through the DE components in the model. In our case, given that by definition $\gamma_\Lambda =0$, we find that $\gamma_{eff} = \gamma_\phi \Omega_\phi /(\Omega_\phi + \Omega_\Lambda)$. 

Clearly, if $\Omega_\phi =0$ ($\Omega_\Lambda =0$), i.e., if $\Lambda$ ($\phi$) is the only DE component then the effective DE EoS simply is $\gamma_{eff} =0$ ($\gamma_{eff} = \gamma_\phi$). More generally, if $\Omega_\phi > 0$ and $\Omega_\phi + \Omega_\Lambda > 0$, a negative value of $\gamma_{eff}$ would indicate a preference of the observations for a phantom-like DE component. This seems to be precisely the case as inferred from the values in Table~\ref{tab:params}: quite consistently $\gamma_{eff} < 0$ at $1-\sigma$. Moreover, the value of the effective DE EoS is practically the same in the presence of the phantom component, irrespective of the model and the form and combination of the DE components, and just a little bit below the phantom divide: $\gamma_{eff} \simeq -0.045$.

\begin{figure}[htp!]
\includegraphics[width=0.49\textwidth]{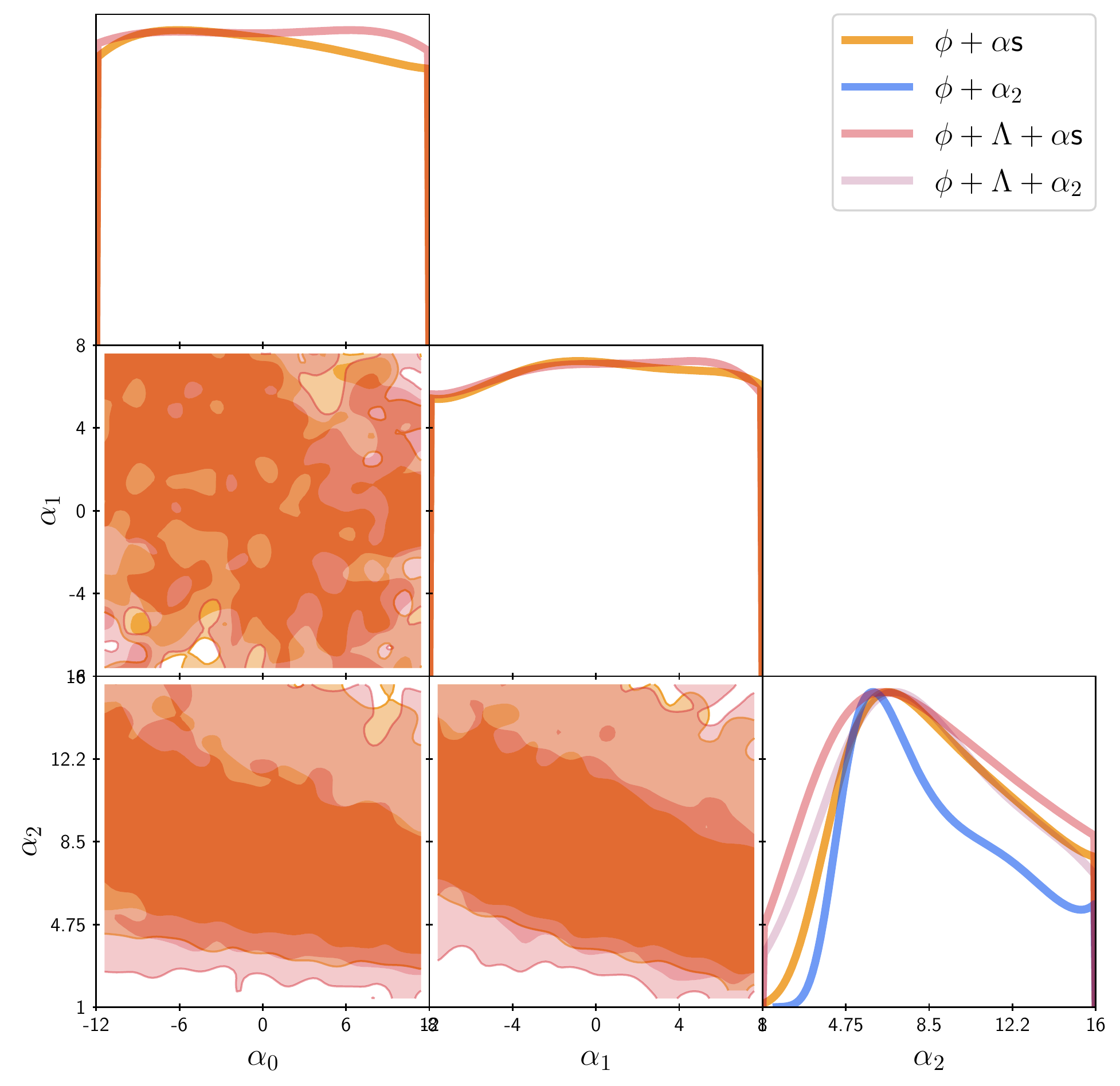}
\caption{\label{fig:alphas} Observational constraints on the active parameters of the phantom potential  $\alpha_0$, $\alpha_1$ and $\alpha_2$. The labels are the same as in Fig.~\ref{fig:general} and Table~\ref{tab:params}. See the text for more details.}
\end{figure}

In Fig.~\ref{fig:alphas} we show the constraints on the active parameters $\alpha$ of the phantom potential, see Eq.~\eqref{eq:GP1}. The overall result is that, independently of the DE model with the phantom field and $\Lambda$, the values of $\alpha_0$ and $\alpha_1$ are completely unconstrained, which means that their inclusion does not make any difference in the fitting to the data, and the latter does not seem to support any added complexity on the phantom models.

Another consequence of the unconstrained values of $\alpha_0$ and $\alpha_1$ is that the ultimate fate of the Universe under the phantom models remains unknown, as any of the late-time values of the EoS discussed in Sec.~\ref{sec:phantom-dominated} is equally likely. In general, the big or little rip solutions cannot be discarded under the models studied here.

Interestingly, the active parameter $\alpha_2$, which also controls the tracker properties of the phantom model, appears to be constrained by the data at around $\alpha_2 \simeq 8.7$. This suggests that the tracker values of the phantom EoS are $\gamma_{\phi,c} \simeq -0.077$ ($\gamma_{\phi,c} \simeq -0.056$) during the radiation (matter) domination era. Although the deviation from the phantom divide is small, it remains to be studied why the data seem to prefer such negative values at early times.

To assess whether the observations have a preference for any of the model variations studied here, we first compute for each one the difference in the value of $\chi^2_{min}$ with respect to $\Lambda$, which curiously enough is the same for all models with phantom: $\Delta \chi^2_{min} = \chi^2_\phi - \chi^2_\Lambda = -5$. This indicates that the quality of the fit increases a bit with the inclusion of $\phi$, irrespective of the presence of $\Lambda$ and of the active parameters $\alpha$'s.

Although there are more free parameters in the phantom models than in the standard $\Lambda$ case (see the number of extra parameters $k$ in Table~\ref{tab:params}) this does not mean that such more general models should be discarded. From a strict Bayesian point of view, for a proper judgment, one must take into account whether the data is able to constrain the extra parameters. This is the case of the active parameters $\alpha_0$ and $\alpha_1$: being unconstrained by the data, the latter does not provide evidence in favour or against the models containing them~\cite{Trotta2008}.

To have a more Bayesian assessment, we also show in Table~\ref{tab:params} the Bayes factors of the phantom models with respect to the $\Lambda$ case, such that $\ln B_{\phi \Lambda} = \ln \mathcal{Z}_\phi - \ln \mathcal{Z}_\Lambda$, where $\mathcal{Z}$ represents the Bayesian evidence. For the calculation of the latter, for each model, we relied on the code \textsc{MCEvidence}~\cite{Heavens:2017afc,Heavens:2017hkr}, which only requires the chains we generated with \textsc{Monte Python}. We see that consistently $\ln B_{\phi \Lambda} > 2$, which means that there is Definite/Positive evidence, under the considered set of observations, in favor of the presence of a phantom DE component.

\subsection{Model selection: Phantom vs $\Lambda$}
Another question that we are interested in is whether data indicates any joint contribution from both the phantom and $\Lambda$ components. To try an answer, we take advantage of the above fact that two of the active parameters are unconstrained and then focus on the models with $\alpha_0 =0 =\alpha_1$, which in turn makes it easier to find the numerical solutions of the phantom models. 

\begin{figure}[htp!]
\includegraphics[width=0.49\textwidth]{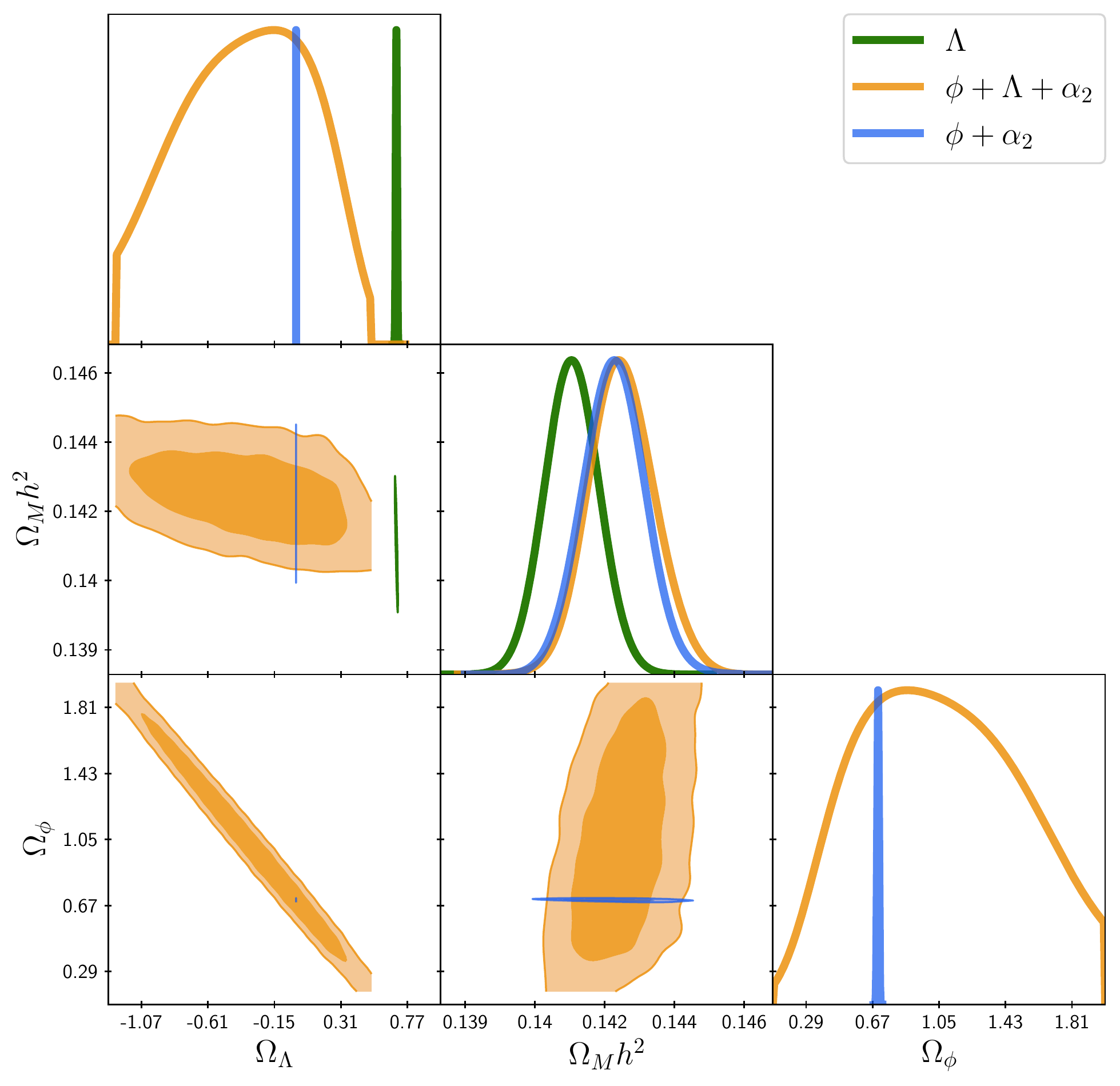}
\caption{\label{fig:DELambda} Observational constraints on $\Omega_\Lambda$, $\Omega_M h^2$, and $\Omega_\phi$ for the extended models indicated in the labels, see also Table~\ref{tab:params}. See the text for more details.}
\end{figure}

The results are shown in Fig.~\ref{fig:DELambda}, for the parameters $\Omega_\Lambda$, $\Omega_M h^2$ and $\Omega_\phi$. The variation in the phantom component $\Omega_\phi$ was extended to the range $[0.1:2]$, with the contribution of $\Omega_\Lambda$ inferred from the Friedmann constraint. This case is called as extended-$\phi + \Lambda +\alpha_2$ in Table~\ref{tab:params}.

The interesting case is the combined presence of the phantom field $\phi$ and $\Lambda$ as DE components (blue contours): the confidence regions seem to suggest a preference for a lower value of $\Omega_\Lambda$, even a negative one. In contrast, there is a preference for large values of the phantom contribution, this time of the order of unity for the density parameter, $\Omega_\phi \simeq 1$. 

However, probably more interesting is that the value inferred for the $\Lambda$-only case (green contour) appears to be located in a low likelihood region when compared with the results of the combination $\phi + \Lambda +\alpha_2$ (orange contour). Correspondingly, the result for $\Omega_\phi$ of the phantom-only case (blue contour) is located within the region of maximum likelihood suggested by the extended case $\phi+\Lambda + \alpha_2$. As seen from Table~\ref{tab:params}, the Bayes factor with respect to the model $\Lambda$, $\ln B_{\phi \Lambda} \simeq +2$, again reinforces our previous result that the data favors the presence of a phantom component in the DE budget.

In other words, the conclusions from the Bayes factor appear to be conservative with respect to the parameter estimation shown in Fig.~\ref{fig:DELambda}: even though we were unable to try the null value $\Omega_\phi=0$ because of numerical limitations, such value seems to be ruled out at $95\%$ confidence level. Hence, the moderate rejection of the model $\Lambda$-only comes from the penalisation the Bayes factor puts on the extended model $\phi+\Lambda + \alpha_2$ for using prior values of $\Omega_\phi$ that yield very low likelihood~\cite{Wagenmakers2010,Trotta2008,Nesseris:2012cq,Jenkins:2011va,Efstathiou:2008ed}.

Taken together: the fit improvement, the conservative rejection from the Bayes factor and the informative posteriors in Fig.~\ref{fig:DELambda}, lead us to conclude that the data seem to rule out a significant contribution of a positive $\Lambda$ in our models; rather, the data seem to prefer the phantom-only model. It is still possible to consider a contribution from a negative $\Lambda$, although none of our aforementioned tests, not even together, gives us decisive hints about such possibility.\footnote{In Appendix~\ref{sec:sddr} we revise the odds of the models in terms of the so-called Savage-Dickey density ratio, which illustrates the interplay of the posterior and the prior of $\Omega_\phi$ on the calculation of the Bayes factor in our models. For comparison, we do the same in Appendix~\ref{sec:sddr1} for the case of a fluid model with a constant EoS accompanying $\Lambda$ as a DE component.} 

\section{\label{sec:discussion}Discussion and conclusions}

In this work, we have studied the tracking behavior of the phantom dark energy models and analyzed its dynamics under a general parameterization of the phantom field potential. For that, we defined a new set of hyperbolic polar variables to write down the Klein Gordon equation of the phantom scalar field as a set of an autonomous dynamical system. The influence of the linear density perturbations has been also incorporated in the analysis. The sufficient and necessary condition for the phantom field to have a tracking behavior also involves just one active parameter and can be generalized even including the other active parameters as long as the $\Omega_{\phi c}$ is negligible in the early Universe. 

Apart from the tracking solutions, dynamics of other kinds of solutions such as scaling, and phantom dominated solutions, are also discussed. We find that scaling solutions do not exist for the phantom model, whereas for the phantom-dominated solutions the asymptotic behavior is similar to the cosmological constant for our choice of $\theta \leq 0$. The numerical solutions for a wide range of active parameters have been studied. It is interesting to note that all solutions for each set of active parameters track the background fluid identically until it reaches the deep in matter domination era. The degeneracy of the solutions is broken at the late time and the present value of the DE EoS depends significantly on the choice of the so-called active parameters.

A combination of recent cosmological data has been used to constrain the cosmological parameters. Three different types of models have been presented: cosmological constant ($\Lambda$), phantom DE ($\phi$), and the phantom DE with cosmological constant ($\phi + \Lambda$). The latter two cases were also studied for pure tracking solution ($\alpha_0 = \alpha_1 = 0$) and general tracking solution ($\alpha_0, \alpha_1 \neq 0$). Different as the case for $\alpha_2$, the statistical analysis can not constraint $\alpha_0, \alpha_1$, which suggests that the tracker value of the phantom EoS was slightly lower than the cosmological constant throughout both the matter and radiation dominated era. Although there is a noticeable shift in the central value of $H_0$ due to the presence of a phantom field it can not solve the $H_0$ tension completely. 

While doing the model comparison using up the concept of the Bayes factor, we found that data favor the existence of phantom DE over the positive cosmological constant. The main result is that a negative cosmological constant can not be ruled out while there is a phantom scalar field component, which agrees with the results obtained in~\cite{Visinelli:2019qqu}.  This may indicate that the dynamics of the DE sector might be more complex than in single-component models. It will be interesting to investigate multicomponent DE models with at least one phantom scalar field, which we expect to present elsewhere.

\begin{acknowledgments}
FXLC acknowledges the receipt of the grant from the Abdus Salam International Centre for Theoretical Physics, Trieste, Italy. This work was partially supported by Programa para el Desarrollo Profesional Docente; Direcci\'on de Apoyo a la Investigaci\'on y al Posgrado, Universidad de Guanajuato; CONACyT M\'exico under Grants No. A1-S-17899, No. 286897, No. 297771, No. 304001; and the Instituto Avanzado de Cosmolog\'ia Collaboration. We acknowledge the use of the Chalawan High Performance Computing cluster, operated and maintained by the National Astronomical Research Institute of Thailand (NARIT); of the COUGHs server at the Universidad de Guanajuato; and the computing facilities at the Laboratorio de Inteligencia Artificial y Superc\'omputo, IFM-UMSNH.
\end{acknowledgments}

\appendix

\section{Phantom perturbations}\label{pert_var}
With the aim of working within the same scheme we used for the background in Section~\ref{sec:mathematical}, where we were able to write down a dynamical system for the KG equation, we now propose the following new variables for the scalar field perturbation $\varphi$ and its derivative $\dot{\varphi}$,
\begin{subequations}
\begin{eqnarray}
\sqrt{\frac{2}{3}} \frac{\kappa \dot{\varphi}}{H} &=& -\Omega^{1/2}_{\phi}e^{\alpha}\cosh(\vartheta/2) \, , \label{eq:22a} \\
\frac{\kappa y_1 \varphi}{\sqrt{6}} &=& -\Omega^{1/2}_{\phi}e^{\alpha}\sinh(\vartheta/2) \, .
\end{eqnarray} 
\end{subequations}

After some algebraic procedure, the equations of motion of linear perturbations~\eqref{eq:perts} can be written in terms of the polar variables $\alpha, \vartheta$ as
\begin{subequations}
\label{eq:pertvardiffeqold}
\begin{eqnarray}
\vartheta' &=& 3\sinh \vartheta -2\frac{k^2}{k_J^2} \left( 1-\cosh \vartheta \right)+y_1\nonumber \\
&& -2e^{-\alpha}h^{\prime}\sinh \left( \frac{\theta}{2} \right)\sinh \left( \frac{\vartheta}{2} \right) \nonumber \\
&&  + \Omega_\phi^{1/2}\left[ \cosh\left( \vartheta + \frac{\theta}{2} \right) - \cosh\left( \frac{\theta}{2} \right) \right]\frac{y_2}{y_1} \, , \nonumber \label{vartheta}\\
\\
\alpha' &=& -\frac{3}{2}\left( \cosh \theta + \cosh \vartheta \right)-\frac{k^2}{k_J^2}\sinh \vartheta  \nonumber \\
&& + e^{-\alpha}h^{\prime}\sinh\left( \frac{\theta}{2} \right)\cosh\left( \frac{\vartheta}{2} \right) \nonumber \\
&& +\frac{\Omega_\phi^{1/2}}{2}\left[\sinh\left( \frac{\theta}{2} \right) - \sinh\left(\vartheta +  \frac{\theta}{2} \right)  \right] \frac{y_2}{y_1}\, .\nonumber \\
\end{eqnarray}
\end{subequations}

If we now define $\delta_0=e^{\alpha}\sinh(\theta/2 + \vartheta/2)$ and $\delta_1=e^{\alpha}\cosh(\theta/2 + \vartheta/2)$, then we can rewrite Eqs.~\eqref{eq:pertvardiffeqold} in terms of the new variables $\delta_0$ and $\delta_1$ to obtain Eqs.~\eqref{eq:deltas}.

\section{Phantom tracker plus $\Lambda$: a nested model and the Savage-Dickey density ratio \label{sec:sddr}}

We use here a common approximation for nested models, the so-called Savage-Dickey density ratio (SDDR)~\cite{Trotta2008,Mukherjee2006} (see also~\cite{Marin2009,Wagenmakers2010} and references therein for more details) to calculate the Bayes factor between the models $\Lambda$ and $\phi +\Lambda$. We can use this approximation because the model $\Lambda$ is properly nested within the model $\phi +\Lambda$: the former is obtained from the latter if we set $\Omega_\phi =0$ (for more details see Appendix~A in~\cite{Wagenmakers2010}). 

The SDDR in our case is then the ratio of the marginalized posterior of $\Omega_\phi$ to its prior, both evaluated at the point $\Omega_\phi$. That is, given the flat prior on $\Omega_\phi$, say in the range $[\Omega_{\phi 1}:\Omega_{\phi 2}]$, the Bayes factor is then 
\begin{equation}
   \ln B = \ln \left[ P(\Omega_\phi) (\Omega_{\phi 2}-\Omega_{\phi 1}) \right] \, , 
\end{equation}
where $P(\Omega_\phi)$ is the marginalized posterior. This is true irrespective of the values taken by the active parameters $\alpha$.

\begin{figure}[htp!]
\includegraphics[width=0.49\textwidth]{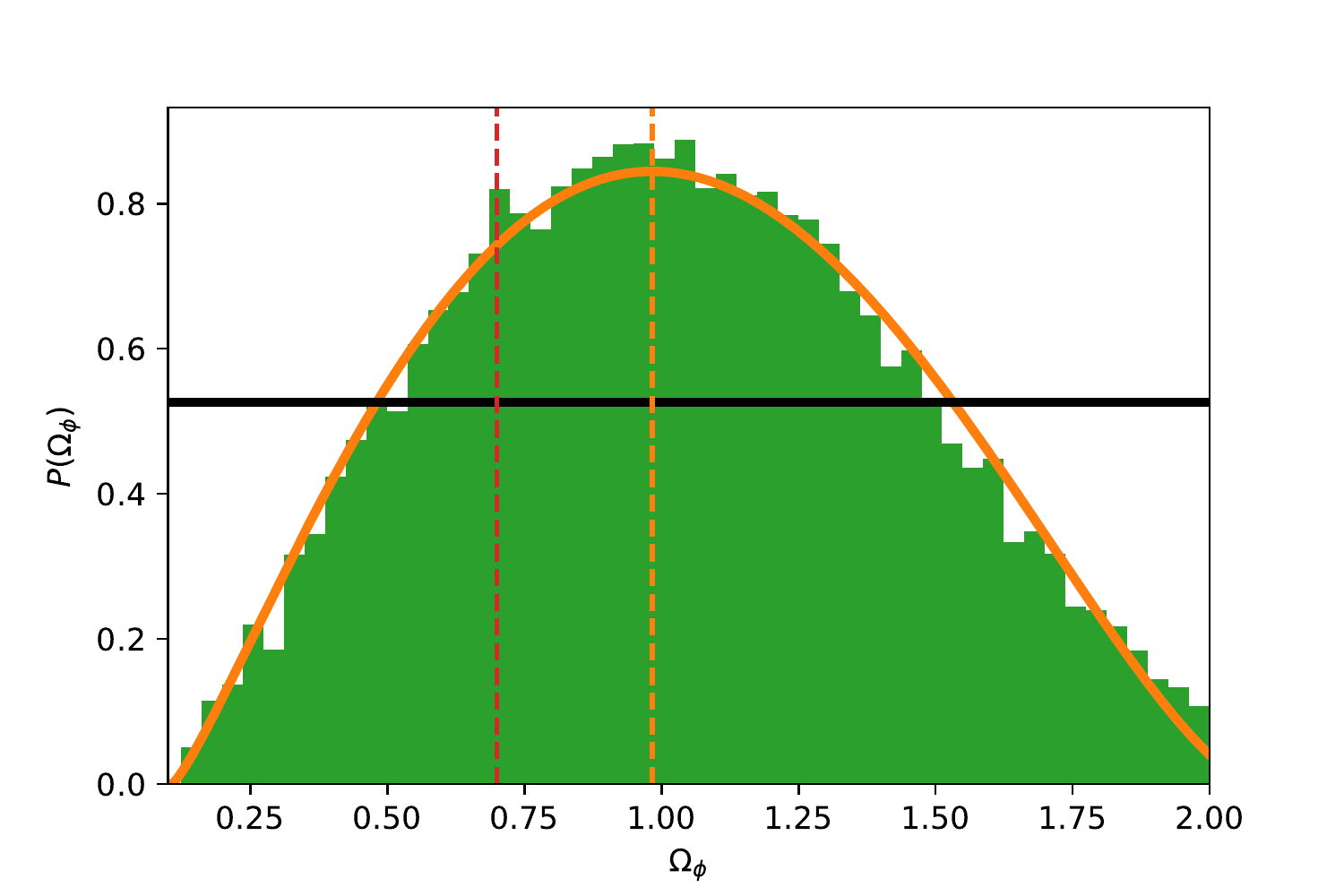}
\caption{\label{fig:histophantom} The (normalized) histogram of the parameter $\Omega_\phi$ obtained for the model $\phi +\Lambda$. The orange curve is a beta PDF fitted to the histogram obtained from the MCMC chains, the horizontal black line represents the flat prior, and the vertical dashed lines indicate the mode of the beta PDF (orange) at $\Omega_\phi = 0.98$ and the value $\Omega_\phi = 0.7$ (red). See the text for more details.}
\end{figure}

In Fig.~\ref{fig:histophantom} we show the marginalized posterior $P(\Omega_\phi)$ for the extended model $\phi +\Lambda + \alpha_2$ after normalization, as calculated from the histogram inferred from the MCMC chains. The orange curve is a Beta PDF fitted to the histogram, whereas the rectangle (black horizontal line) with height $1/1.9$ represents the prior. 

Although we were not able to explore the values $\Omega_\phi < 0.1$ because of numerical limitations, it is clear that our results strongly suggest that $\ln B \to -\infty$ as $\Omega_\phi \to 0$, and in consequence the simplest model $\Lambda$, with no phantom contribution, appears to be strongly rejected by the data.

Another possibility we can explore is to consider a model without $\Lambda$ ($\Omega_\Lambda =0$), which corresponds to the value for which the phantom field $\phi$ makes up the whole of the DE budget at $\Omega_\phi \simeq 0.7$ (vertical dashed red line in Fig.~\ref{fig:histophantom}). For this latter value, the Bayes factor is $\ln B = 0.19$, which means that the evidence is inconclusive for $\Omega_\Lambda =0$. Actually, the mode of the beta PDF in Fig.~\ref{fig:histophantom} is located at $\Omega_\phi \simeq 0.98$, for which we get $\ln B = 0.41$, and then the evidence is also inconclusive with respect a negative value of $\Lambda$ (in this case corresponding to $\Omega_\Lambda \simeq -0.28$).

In summary, the SDDR gives results consistent with our calculations in Sec.~\ref{sec:numerical}, in that there is strong evidence in favor of the presence of a phantom component, but the difference between a purely phantom DE and a combination with a negative $\Lambda$ is not conclusive.

\section{Fluid $F$ plus $\Lambda$ \label{sec:sddr1}}
To compare the results in the main text with another type of DE model, we repeated the calculations for a phantom fluid ($F$) with a constant EoS $w_0$, which is the simplest generalization from a cosmological constant. 

\begin{figure}[htp!]
\includegraphics[width=0.48\textwidth]{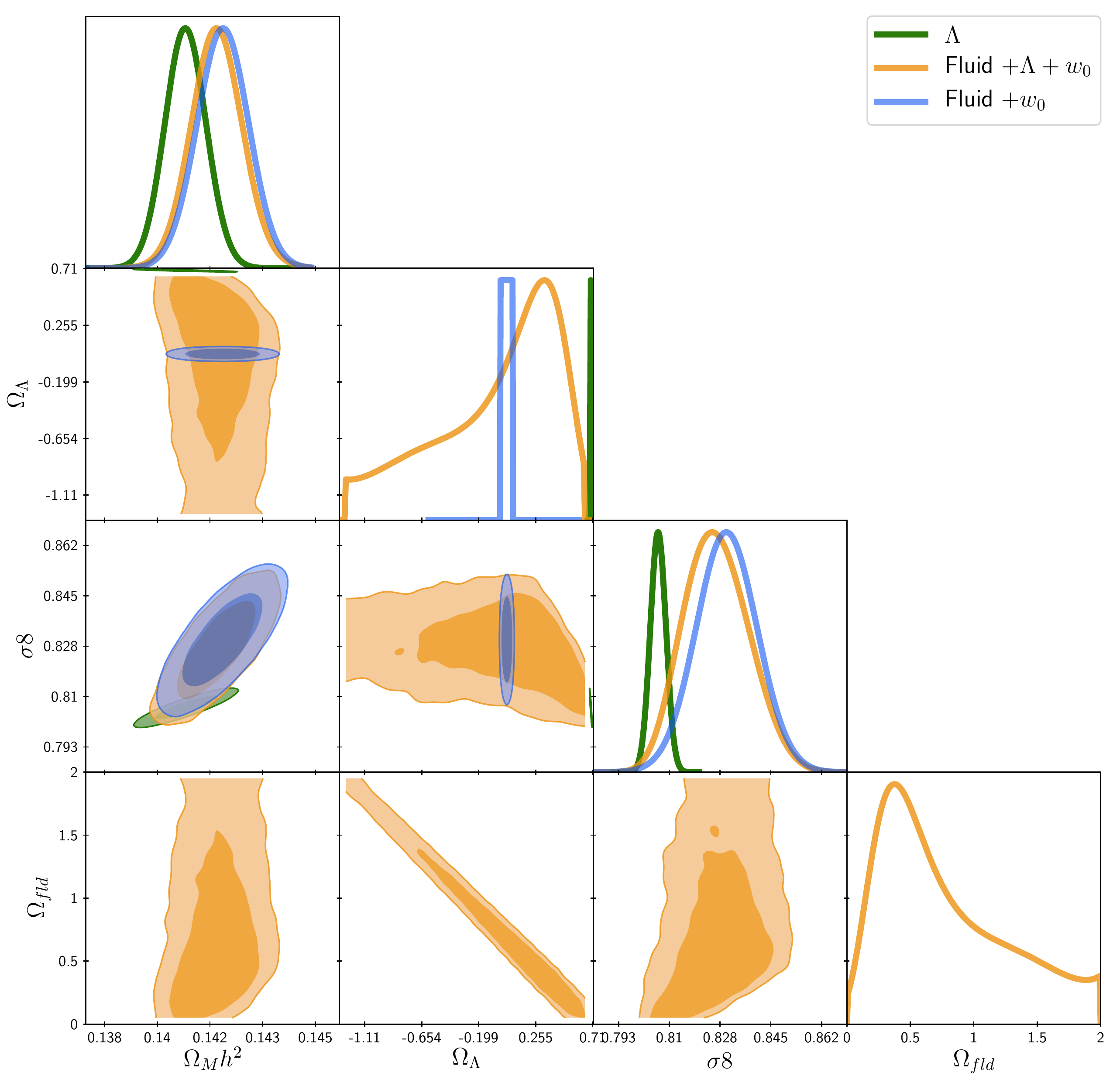}
\includegraphics[width=0.49\textwidth]{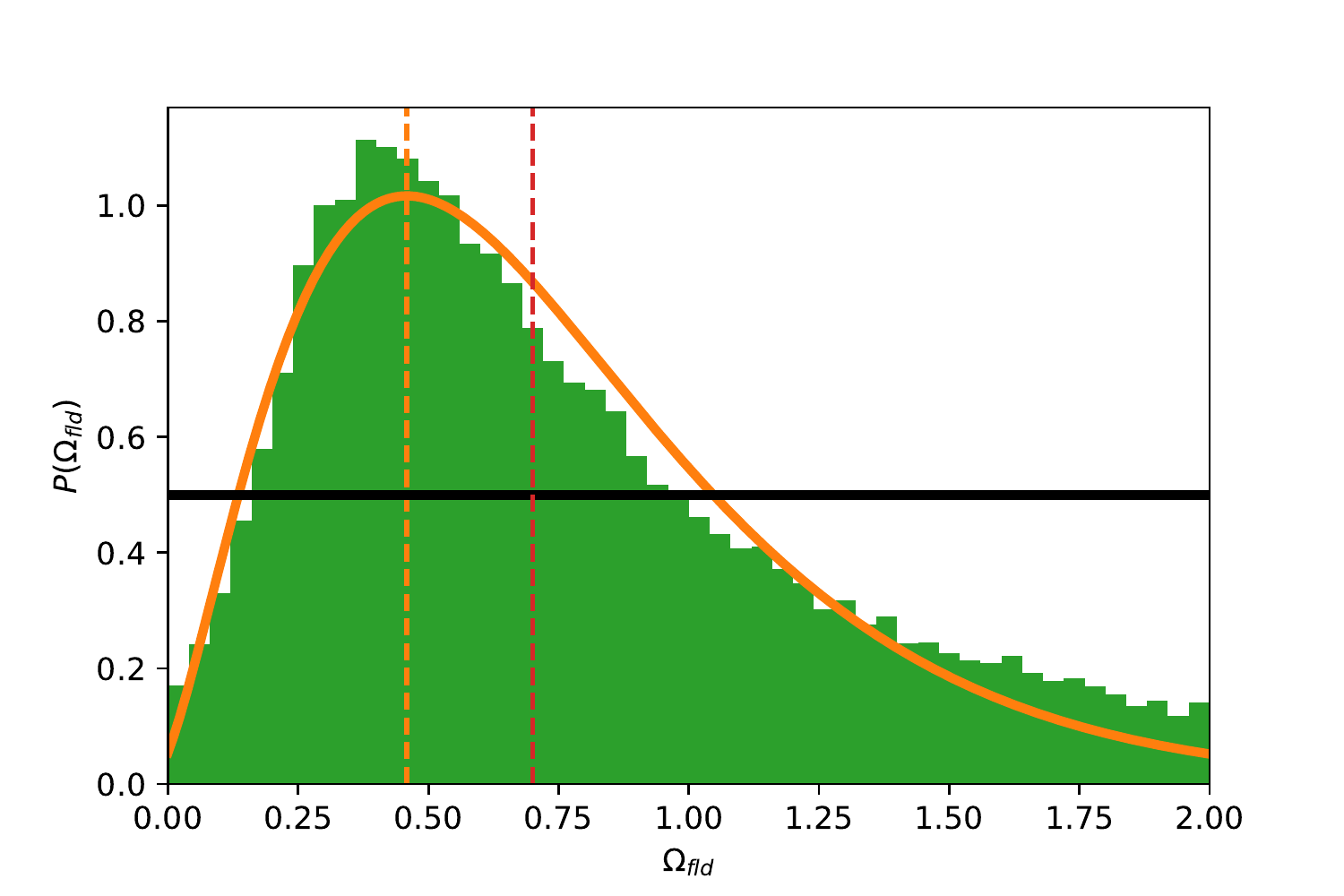}
\caption{\label{fig:DECPL} (Top) Observational constraints on $\Omega_\Lambda$, $\Omega_M h^2$, $\sigma_8$, and $\Omega_{fld}$ for the models with a cosmological constant $\Lambda$, with a combination of a fluid plus a cosmological constant $F+\Lambda$, and with only a fluid component $F$. (Bottom) The (normalized) histogram of the parameter $\Omega_{fld}$ obtained for the model $F +\Lambda$. The orange curve is a Gamma PDF fitted to the histogram obtained from the MCMC chains, the horizontal black line represents the flat prior, and the vertical dashed lines indicate the mode of the Gamma PDF (orange) at $\Omega_{fld} = 0.46$ and the value $\Omega_{fld} = 0.7$ (red). See the text for more details.}
\end{figure}

The DE budget is then composed of a general fluid and $\Lambda$, and we varied the fluid contribution and its EoS in the ranges $\Omega_{fld} = [0:2]$ and $w_0 =[-1.2:-0.8]$. The resultant plots, after the comparison with the same set of data as for the phantom field in the main text are shown in the top panel of Fig.~\ref{fig:DECPL}, whereas the fitted values are listed in Table~\ref{tab:params-cpl} (for comparison see Table~\ref{tab:params}). The Bayes factor were also calculated with the code \textsc{MCEvidence}~\cite{Heavens:2017afc,Heavens:2017hkr}.

\begin{table}[htp!]
\caption{\label{tab:params-cpl} Fitted values of the free parameters in the models with a general fluid. The confidence regions for the parameters are shown in Fig.~\ref{fig:DECPL} (top). See the text for more details.}
\begin{ruledtabular}
\begin{tabular}{|c|c|c|}
Parameter & Fluid$+w_0$ & Fluid$+ \Lambda +w_0$ \\ 
\hline
$H_0$ & $69.3_{-0.67}^{+0.67}$ & $69.14_{-0.73}^{+0.65}$ \\
\hline
$\Omega_M h^2$ & $0.1419_{-0.00090}^{+0.00091}$ & $0.1418_{-0.00089}^{+0.00091}$ \\
\hline
$\gamma_{eff}$ & $-0.0655_{-0.000904}^{+0.000910}$ & $-0.0503_{-0.029}^{+0.029}$ \\
\hline
$\Omega_\Lambda$ & $0$ & $-0.06234_{-0.25}^{+0.64}$ \\
\hline
$\Omega_{fld}$ & $0.703_{-0.0055}^{+0.0058}$ & $0.7643_{-0.65}^{+0.25}$ \\
\hline
$k$ & $+1$ & $+2$ \\
\hline
$\Delta \chi^2_{min}$ & $-5$ & $-5$ \\
\hline
$\ln B_{F \Lambda}$ & $+1.04$ & $+0.79$ \\
 & Weak & Inconclusive
\end{tabular}
\end{ruledtabular}
\end{table}

The fit to the data is again improved with respect to the $\Lambda$ only case, and the results on the different observables look quite similar to those obtained for the phantom field (see for instance Fig.~\ref{fig:DELambda}). However, the Bayes factors indicate that the evidence in favor of the presence of the fluid component is at most weak with respect to $\Lambda$ only. 

This can be verified also by means of the SDDR as in Appendix~\ref{sec:sddr} above, and then the Bayes factor can be written as,
\begin{equation}
   \ln B = \ln \left[ P(\Omega_{fld}) (\Omega_{fld 2}-\Omega_{fld 1}) \right] \, , \label{eq:sddr-cpl}
\end{equation}
where $P(\Omega_{fld})$ is the marginalized posterior of $\Omega_{fld}$, the latter represented by the histogram shown in the bottom panel of Fig.~\ref{fig:DECPL}. The orange curve is a Gamma PDF fitted to the histogram, whereas the rectangle (black horizontal line) with height $0.5$ represents the prior.

The Bayes factor for $\Omega_{fld}=0$, according to Eq.~\eqref{eq:sddr-cpl}, is $\ln B = -2.21$, whereas for $\Omega_\Lambda=0$, with only the fluid component as DE, is $\ln B = 0.55$. Moreover, the mode of the Gamma PDF is located at $\Omega_{fld} \simeq 0.46$, with corresponding Bayes factor $\ln B = 0.71$. In overall, these results suggest that the most likely scenario resembles more the equipartition of the DE budget between $\Lambda$ and the fluid component, where $\Lambda$ remains positive definite (see also~\cite{Calderon:2020hoc} for a similar study but different results).

\bibliography{apssamp}

\end{document}